\documentclass[12pt]{article}
\usepackage{cite,epsfig,amssymb,amsmath,amstext,  graphicx,color,subfigure,xcolor,mathrsfs,tikz}
\evensidemargin \oddsidemargin

\numberwithin{equation}{section}

\usetikzlibrary{decorations.pathmorphing}
\usepackage{latexsym}
\usepackage{hyperref}
\usepackage{bbm}
\usepackage{caption}
\hypersetup{colorlinks,%
  citecolor=blue,%
  linkcolor=red,%
  pdftex}
\usepackage[numbers,sort&compress,square]{natbib}
\usepackage{varioref}
\usepackage{slashed}
\usepackage[normalem]{ulem}
\usepackage[makeroom]{cancel}
\usepackage{pdfsync}
\usepackage{mathrsfs}
\begin{document}

~
\vspace{5mm}

\begin{center}

{{\Large \bf Supersymmetric Backgrounds in $(1+1)$ Dimensions and Inhomogeneous Field Theory}}
\\[5mm]

Jeongwon Ho$^1$, ~~O-Kab Kwon$^2$, ~~Sang-A Park$^{2,3}$, ~~Sang-Heon Yi$^1$  \\[2mm]
{\it $^1$Center for Quantum Spacetime, Sogang University, Seoul 04107, Republic of Korea} \\
{\it $^2$Department of Physics,~BK21 Physics Research Division,
 Autonomous Institute of Natural Science,~Institute of Basic Science, Sungkyunkwan University, Suwon 16419, Korea} \\
{\it $^3$Department of Physics, Yonsei University, Seoul 03722, Korea} 
\\[2mm]
{\it freejwho@gmail.com, ~okab@skku.edu, ~psang314@gmail.com, ~shyi@sogang.ac.kr}
\end{center}
\vspace{15mm}

\begin{abstract}
\noindent
We find a $(1+1)$-dimensional metric solution for a background hosting various supersymmetric field theories with a single non-chiral real supercharge. This supersymmetric background is globally hyperbolic even though it contains a naked null singularity. In this regard, we show that scalar wave propagation on the background is well-defined and so the curvature singularity is a {\it mild} one. Taking inspiration from our previous work, we relate the field theory on this curved background to some classes of $(1+1)$-dimensional inhomogeneous field theory in the supersymmetric setup. Utilizing our supersymmetric background, we elucidate the limitations of canonical quantization and highlight the conceptual advantages of the algebraic approach to quantization.
\end{abstract}


\vskip 1.0in

\hrulefill

All authors contributed equally.

\thispagestyle{empty}
 
\newpage
{
  \hypersetup{hidelinks}
  \tableofcontents
}

\section{Introduction}
Recently, field theory models with spacetime dependent mass and couplings have been actively investigated. For instance, physical systems having spacetime dependent masses  were studied in various research areas, such as  condensed matter physics~\cite{Bastard:1988, Roos:1983, Filho:2011}, neutron physics~\cite{Ring:1980, Chamel:2005nd}, and cosmology~\cite{Linde:1993cn}.

In the context of string theory, mass and coupling parameters in low energy effective field theories can be understood as non-dynamical  traces  (or backgrounds) of high energy fields. Usually, these mass and coupling parameters are taken as constants in order to maintain Poincar\'{e} symmetry of theories. For instance,   gauge coupling in $\mathcal{N} = 4$ super Yang-Mills theory is dual to  dilaton field of type IIB supergravity, and  it is well known that the constant gauge coupling is the  constant boundary value of dilaton field.  As additional interesting examples, one can recall  mass parameters in $\mathcal{N}=1^*$ (or $\mathcal{N} = 2^*$)  super Yang-Mills theory and in $\mathcal{N} = 6$  mass-deformed ABJM theory~\cite{Aharony:2008ug, Hosomichi:2008jb}, which are traces of constant RR 7-form field strength in type IIB supergravity and of constant 4-form field strength in 11-dimensional supergravity, respectively. However, in general, these non-dynamical traces become {\it spacetime dependent} mass and coupling parameters, resulting in some or all of the Poincar\'{e} symmetry breaking.  Such field theory with {\it spacetime dependent} mass and coupling parameters will be termed as inhomogeneous field theory (IFT) in the following. Since supersymmetry is tied with Poincar\'{e}  symmetry, it is broken as the Poincar\'{e} symmetry is. Consequently, the number of supersymmetries in IFTs could be reduced compared to that of homogeneous field theory.

 This extension to IFT has been made in various directions: Janus Yang-Mills theory in 4-dimensions, which is dual to gravity with position-dependent dilaton deformation  in  AdS$_5$,  has position-dependent gauge coupling~\cite{Bak:2003jk}. This theory has been supersymmetrized with a reduced number of supercharges~\cite{DHoker:2006qeo, Kim:2008dj, Kim:2009wv}. $\mathcal{N}=4$ super Yang-Mills theory and  $\mathcal{N} = 6$  ABJM theory are extended in such a way that   mass parameters are position-dependent and the number of supercharges reduces as well~\cite{Kim:2018qle,Kim:2019kns, Arav:2020obl,Kim:2020jrs, Gauntlett:2018vhk, Arav:2018njv, Ahn:2019pqy, Hyun:2019juj, Arav:2020asu}.  As a $(1+1)$-dimensional model, the $\mathcal{N}= (1,1)$ Wess-Zumino model is also extended to  a half supersymmetric $\mathcal{N}= (\frac{1}{2},\frac{1}{2})$ model with position-dependent couplings~\cite{Kwon:2021flc}, which will be discussed in detail later.

The broken Poincar\'{e} symmetry related to the reduced supersymmetry causes a difficulty to quantize IFT.   Recently, we proposed the conceptually consistent  method for quantizing IFT, based on the conversion relation between field theory on curved spacetime (FTCS)
 and  IFT in $(1+1)$-dimensions and the algebraic approach\footnote{For review, see \cite{Haag:1992hx,Wald:1995yp,Yngvason:2004uh,Halvorson:2006wj,Hollands:2009bke,Benini:2013fia,Hollands:2014eia,Khavkine:2014mta,Fredenhagen:2014lda}.  } to quantization~\cite{Ho:2022ibc}. In this paper, motivated by  our previous proposal on bosonic scalar field, we explore a possibility of correspondence in a supersymmetric setup. 

Since, on the contrary to the bosonic case,  supersymmetry restricts allowed background spacetimes,  we need to obtain  specific backgrounds on which supersymmetric field theories can reside. In that sense, it is meaningful in itself to find a rigid curved background which we will call supersymmetric background. For instance, Minkowski and AdS spacetimes are well-known examples as such supersymmetric backgrounds.   This situation may be thought somewhat odd since numerous curved spacetimes are known in the supergravity context.  A recent work~\cite{Festuccia:2011ws} has filled the gap and provided a new way to construct  supersymmetric backgrounds. While many examples constructed in this way are Euclidean spaces,  in this paper, we present a $(1+1)$-dimensional supersymmetric background  on which various field theories can live.

To construct a supersymmetric field theory on curved spacetime (SFTCS), we use a method that differs slightly from the standard steps developed for supersymmetrizing a Lagrangian in \cite{Festuccia:2011ws}. 
This procedure gives us a kind of  Killing spinor equation. After  solving the Killing spinor equation, we check that Minkowski/Rindler and AdS$_{2}$ spaces are supersymmetric backgrounds. In addition, we find a supersymmetric background with naked null singularity. It is shown that the singularity is {\it mild} in the context of the scalar wave propagation on that background. To show the well-posedness of scalar wave propagation as an initial value problem, we investigate the self-adjointness of a relevant operator~\cite{Reed}. Interestingly, the usual test for the essential self-adjointness of a symmetric operator is greatly simplified by the fact that the symmetric operator can be understood as a hamiltonian of supersymmetric quantum mechanics (SQM)~\cite{Cooper:1994eh}.   In fact, we show that the equation of the bosonic part ({\it i.e.} scalar field)  of free SIFT can be rewritten as a SQM.

Taking inspiration from the bosonic correspondence proposed in \cite{Ho:2022ibc}, we are able to slightly generalize SIFT models and to find a novel supersymmetric background in the context of SFTCS. Despite the seemingly specific nature of our supersymmetric background, we are able to obtain complete mode solutions to free scalar field equation, and then canonically quantize the scalar field. It turns out that the canonical quantization seems obscured to interpret our system. Resultantly, we affirm that the algebraic approach gives comprehensive method for quantization in IFT.

Our paper is organized as follows. In Section \ref{SFTCS}, we  present a  specific procedure to construct SFTCS. As an interesting example, we obtain a supersymmetric background with naked null singularity. In Section \ref{SIFT}, we extend SIFT with spacetime-dependent mass and couplings from the position-dependent ones. It turns out that the SUSY variation parameters need to be elevated to spacetime-dependent ones. In Section \ref{wavedy}, we solve the scalar wave equation on our supersymmetric background and show  the well-posedness of the scalar wave propagation,   concluding the {\it mildness} of our background singularity. 
We study scalar field quantization in the absence of preferred vacuum and find an algebraic description to be more appropriate for consistent interpretation.
In the final section, we summarize our results and discuss  future research directions.
In Appendix, we present notation, relevant solutions, and various useful formulae.

\section{Supersymmetric Field Theory on Curved Spacetime}\label{SFTCS}

In the previous work~\cite{Ho:2022ibc}, we have proposed a quantization method of IFT in (1+1) dimensions based on the `relation' between a specific scalar IFT and a scalar FTCS, and algebraic approach to FTCS.  Those actions are given respectively by  
\begin{align} 
& \!\! \!S_{{\rm IFT}} =\int d^2 x \mathcal{L}_{{\rm IFT}} = \int d^2 x\Big[ -\frac{1}{2}\eta^{\mu\nu}\partial _{\mu}\phi \partial_{\nu}\phi - \frac{1}{2}m^{2}(x)\phi^{2} - \sum_{n=3}g_{n}(x)\phi^{n} +J(x)\phi\Big]\,,
\label{IFTact1}\\
&\!\! \!S_{\rm FTCS} =\int d^2x\sqrt{-g} \, \mathcal{L}_{\rm FTCS} =\int d^2x\sqrt{-g}\, \Big[ -\frac{1}{2}g^{\mu\nu}\nabla_{\mu}\phi\nabla_{\nu}\phi -\frac{1}{2}m_0^2\phi^2 -\sum_{\ell=1} h_{\ell}(\mathcal{R})\phi^{\ell} \Big]\,,\label{FTCSact1}
\end{align}
where $m^2(x)$ denotes a position-dependent squared mass and $m_0$ is a constant. Here $h_{\ell}(\mathcal{R})$'s are functions of the curvature scalar $\mathcal{R}$, while $g_n(x)$ and $J(x)$ are specific functions of $x$ corresponding to  given  $h_{\ell}(\mathcal{R})$'s in order to guarantee the conversion from \eqref{FTCSact1} to \eqref{IFTact1}.  For more details, see \cite{Ho:2022ibc}.

The    IFT action \eqref{IFTact1} was  supersymmetrized in \cite{Kwon:2021flc}.  This supersymmetrized  action has {\it  only one real supercharge},  the number of which  is half the number of supercharges of the $\mathcal{N} = (1,1)$ Wess-Zumino model in $(1+1)$-dimensions.
This $\mathcal{N}=(1,1)$ model has constant coupling parameters, unlike   the supersymmetric version of \eqref{IFTact1}. The supersymmetrized model in \cite{Kwon:2021flc} is  different from the $\mathcal{N}=(1,0)$  model~\cite{Sakamoto:1984zk, Hull:1985jv, Brooks:1986uh}, which also has one real supercharge.  To distinguish from other cases, we refer the model in \cite{Kwon:2021flc} as $\mathcal{N}=(\frac{1}{2}, \frac{1}{2})$ supersymmetric one.   

Considering  the relation between the two bosonic actions,  \eqref{IFTact1} and \eqref{FTCSact1}, in \cite{Ho:2022ibc}, we anticipate the supersymmetrized version of the relation. Inspired by this, we expect that the bosonic action \eqref{FTCSact1} would be supersymmetrized as well. However, as was pointed out in  \cite{Festuccia:2011ws}, one cannot construct supersymmetric field theory on an arbitrary curved background. In this section, we  find a specific background geometry for the supersymmetrization of the action \eqref{FTCSact1} {\it with one real supercharge}.

In \cite{Festuccia:2011ws}, the authors  have developed a systematic method for constructing supersymmetric field theories on specific curved backgrounds.
In this method, one starts with given supergravity and matter multiplets  in the limit of the Planck mass  $M_{{\rm Pl}} \to \infty $  with  appropriate scaling of the various auxiliary fields in the supergravity multiplet. In this so-called rigid limit, the gravity multiplet becomes non-dynamical, whereas imposing certain conditions, matter fields remain dynamical and have  global supersymmetry.
The vanishing of the supersymmetry variation of the gravitino  determines the allowed bosonic part of non-dynamical gravity multiplet, which are essentially given by the `generalized' Killing spinor equation in the form of
\begin{align} \label{SKE}
\nabla_\mu \epsilon(\pmb{x}) = M_\mu (\pmb{x}) \epsilon(\pmb{x}),
\end{align}
where $\nabla_\mu$ is the covariant derivative of spinor defined in Appendix \ref{AppA} and $\pmb{x}$ denotes a spacetime  point, $\pmb{x} = (t,x)$ in our case.  Here, $\epsilon$ denotes a supersymmetric parameter and $M_\mu (\pmb{x})$ is written in terms of the bosonic part of the non-dynamical gravity multiplet with spinor indices.
Consequently, a rich landscape for rigid supersymmetric field theory models in curved spacetime was uncovered. 
On the other hand, it is not straightforward to apply this method directly to our case.
In the following, we take an alternative route for supersymmetrization under the same spirit in \cite{Hama:2011ea, Festuccia:2011ws}.

First of all, let us recall the supersymmetrization of a real scalar field theory in  Minkowski background. It is well-known that the  $\mathcal{N}=(1,1)$ Wess-Zumino  action  is given by 
\begin{align}\label{SFT1}
S_{\rm SFT} =\int d^2 x\mathcal{L}_{\rm SFT} = \int d^2 x \Big[ -\frac{1}{2} \eta^{\mu\nu}\partial_{\mu} \phi \partial_{\nu} \phi + \frac{i}{2} \bar \psi \gamma^{\mu}_{\rm F} \partial_{\mu} \psi + \frac{i}{2} \left( \frac{\partial^2 W}{\partial \phi^2} \right) \bar \psi \psi - \frac{1}{2}  \left(\frac{\partial W}{  \partial\phi}\right)^2  \Big],
\end{align}
where $x^{\mu} = t,x$ denote the flat coordinates and $\gamma^{\mu}_{\rm F}$ are gamma matrices in flat spacetime. The superpotential $W(\phi)$ has the following form 
\begin{align}\label{W(phi)}
W(\phi)  = \sum_{n\ge2} \lambda_n  \phi^n
\end{align}
with constant couplings $\lambda_n$'s.

\subsection{Model construction}

To construct a supersymmetric field theory on a curved spacetime background\footnote{ At this stage,  we consider an arbitrary  background. }, we simply replace the flat metric in \eqref{SFT1} by a curved one and obtain
\begin{align}\label{act0}
        S_{{\rm SFT}}^{g}= \int d^2 x \sqrt{-g}\, \mathcal{L}_0 
\end{align}
with 
\begin{align}\label{L0}
\mathcal{L}_0 = -\frac{1}{2}g^{\mu\nu}\nabla_{\mu}\phi\nabla_{\nu}\phi +\frac{i}{2}\bar{\Psi}\gamma^{\mu}\nabla_{\mu}\Psi +\frac{i}{2}\left( \frac{\partial^2 W}{\partial \phi^2} \right)\bar{\Psi}\Psi -\frac{1}{2}\left(\frac{\partial W}{  \partial\phi}\right)^2\, ,
\end{align}
where $\mu, \nu $ are curved spacetime indices and $\Psi$ denotes the fermion field in a curved spacetime.
The convention for fermion fields and gamma matrices in curved spacetime is summarized in Appendix \ref{AppA}.
However,  the action \eqref{act0} is not supersymmetric, in general.  As a method of supersymmetrization of the action \eqref{act0} preserving   covariance, we extend the superpotential \eqref{W(phi)} as
\begin{align}\label{spot2}
        W(\phi)\,\, \longrightarrow\,\,  \mathcal{W}(\phi,\mathcal{R})=\sum_{n  \ge 1} \mathcal{F}_{n} (\mathcal{R})\phi^n\,,
\end{align}
where $\mathcal{F}_n$'s are some functions of curvature scalar $\mathcal{R}$. Such an extension is motivated by the relation between bosonic actions in \eqref{IFTact1} and \eqref{FTCSact1}~\cite{Ho:2022ibc}. 

 Under  the supersymmetry  variation,
\begin{align}\label{svar1}
        \delta \phi &= i\bar{\Psi}\epsilon\,, \nonumber \\
        \delta \Psi &= -\gamma^{\mu}\nabla_{\mu}\phi~ \epsilon + \Big( \frac{\partial \mathcal{W} }{\partial \phi}\Big) \epsilon\,,
\end{align}
the variation of the Lagrangian in \eqref{act0} results in 
\begin{align}\label{varL}
        \delta(\sqrt{-g}\mathcal{L}_0) &= -i \sqrt{-g}~\nabla_{\mu}\phi\, \bar{\Psi}(g^{\mu\nu}-\gamma^{\mu\nu})\nabla_{\nu}\epsilon 
 +i\sqrt{-g}\sum_{n} n\phi^{n-1}\bar{\Psi}\gamma^{\mu} \nabla_{\mu} \big(\mathcal{F}_n\epsilon \big),
\end{align}
where  $\gamma^{\mu \nu} \equiv  \frac{1}{2} \left( \gamma^\mu \gamma^\nu - \gamma^\nu \gamma^\mu \right)$.
Here, the supersymmetric parameter  $\epsilon$ is a two-component Majorana spinor.

We can consider the most general form of $M_{\mu}(\pmb{x})$ in \eqref{SKE}
as
\begin{align}\label{genKS}
        \nabla_{\mu}\epsilon =M_\mu \epsilon =b_\mu \mathbbm{\epsilon} + b_{\mu \nu} \gamma^\nu \epsilon + b_{\mu\rho\sigma} \gamma^{\rho\sigma} \epsilon, \qquad  b_{\mu \rho\sigma} = - b_{\mu \sigma\rho},
\end{align}
where $b_\mu, b_{\mu\nu}$, and $b_{\mu\nu\rho}$ are functions of background metric\footnote{Since a superfield formalism~\cite{Festuccia:2011ws} is not adopted in this paper,  auxiliary fields do not appear in the $b $-tensors.}. To remove  the first term in the right-hand side of \eqref{varL}, one can take $b_{\mu\rho\sigma} \gamma^{\rho\sigma} = b^\nu \gamma_{\mu \nu}$ and $b_{\mu \nu} + b_{\nu \mu} = g_{\mu \nu} b^\rho_{~\rho} $. Let us introduce $f$ as
\begin{align}
        f\equiv b^{\mu}{}_{\mu}-b_{\mu\nu}\gamma^{\mu\nu}\,,
\end{align}
which is taken in the form  of $f = f(\mathcal{R})$ for general covariance. Although $f$ may be a function of the derivative of $\mathcal{R}$ and terms of auxiliary fields in superfield formalism as well as $\mathcal{R}$, we take $f$ as a function of $\mathcal{R}$ only for the fulfillment of our purpose and simplicity. Here, we consider only the term proportional to $\gamma_{\mu}$ for simplicity in $M_{\mu}$. Then, \eqref{genKS} reduces to
\begin{align}\label{mainKS}
\nabla_{\mu}\epsilon  =\frac{1}{2}f\gamma_{\mu}\epsilon.
\end{align}
In the following, we call this equation as the `generalized' Killing spinor equation. 
Taking a covariant derivative on both sides of the above equation, we obtain \begin{align}\label{mainKS2}
\frac{1}{2}(\mathcal{R}+2f^2) \epsilon=-\nabla_{\mu}f\gamma^{\mu}\epsilon.
\end{align} 
Requiring an additional condition 
\begin{align} \label{cond3}
        \nabla_{\mu}\mathcal{F}_{n}(\mathcal{R})\gamma^{\mu}\epsilon=\mathcal{G}_{n}(\mathcal{R})\epsilon\,,
\end{align}
together with the condition \eqref{mainKS}, we obtain the following supersymmetric Lagrangian 
\begin{align}\label{L2}
        \sqrt{-g}\, \mathcal{L}_{\rm SFTCS}= \sqrt{-g} \big(\mathcal{L}_0 - f(\mathcal{R})\mathcal{W}(\phi, \mathcal{R}) - \mathcal{U}(\phi, \mathcal{R}) \big), 
\end{align}
where 
\begin{align}\label{U}
\mathcal{U}(\phi, \mathcal{R})\equiv \sum_{n\geq1} \mathcal{G}_n(\mathcal{R})\phi^n.
\end{align}
It has to be emphasized that the additional requirement \eqref{cond3} on the Killing spinor, $\epsilon$, reduces the number of supersymmetries ({\it i.e.} the number of independent parameters in the solution of $\epsilon$) of the above model. On the contrary, in the case of $\mathcal{G}_{n}(\mathcal{R})=0$, or equivalently $\mathcal{F}_n(\mathcal{R})=constant$, the number of supersymmetries is determined by \eqref{mainKS} only.

By solving both \eqref{mainKS} and \eqref{cond3} simultaneously, we obtain $\mathcal{G}_n(\mathcal{R})$
as 
\begin{align}\label{G_n}
        \mathcal{G}_{n} = - \frac{1}{2}(\mathcal{R}+2f^2)\frac{\left(\frac{\partial \mathcal{F}_n}{\partial \mathcal{R}} \right)}{\left(\frac{\partial f}{\partial \mathcal{R}} \right)}, 
\end{align}
where we have assumed $f (\mathcal{R}) \ne {\it constant} $\footnote{This means that there exists only one supersymmetry parameter corresponding to one real supercharge.  }.
The above expression can also be rewritten as
\begin{align}\label{G}
\sqrt{-g}\, \mathcal{G}_{n}{}^2=-4\partial_{+}\mathcal{F}_n\partial_{-}\mathcal{F}_n
\,.
\end{align}
In the case of $f (\mathcal{R}) = {\it constant}$, which implies $\mathcal{R}={\it constant}$, $\mathcal{F}_n (\mathcal{R})$'s become constant functions and so $\mathcal{G}_n(\mathcal{R})=0$  from \eqref{cond3}. Note that this functional form of $\mathcal{F}_n (\mathcal{R})$ corresponds to  $\mathcal{W}(\phi, \mathcal{R} = constant) = W(\phi)$ in \eqref{W(phi)}.

In fact, \eqref{cond3} does not impose additional restriction except for  \eqref{G_n}. This means that arbitrary forms of $\mathcal{F}_{n}(\mathcal{R})$ are allowed  as far as the background satisfies the condition \eqref{mainKS}. Therefore, we will consider only the condition \eqref{mainKS} in the following. Now two cases are investigated separately: $f (\mathcal{R}) = {\it constant} $ and  $f (\mathcal{R}) \ne {\it constant} $. Before going ahead, we recall that, in the conformal gauge, the metric in $(1+1)$ dimensions is generically given by   
\begin{align}\label{congau}
ds^2 = e^{2 \omega (t,x)} \left( -dt^2 + dx^2 \right)\, . 
\end{align}
Now, we adopt this general metric in solving the Killing spinor equation \eqref{mainKS} with the condition \eqref{cond3}.

\subsection{  $f (\mathcal{R}) = f_0 = {\it constant}$ }

 In this subsection, we set $f (\mathcal{R}) = f_0$, which means $\mathcal{R} ={\it constant}$. From the condition \eqref{cond3}, one can see that the corresponding  supersymmetric Lagrangian is given by \eqref{L2} without the last term,  $\mathcal{U}(\phi, \mathcal{R})$. 

First of all, when $f_0=0$, the solution to the Killing spinor equation \eqref{mainKS} is given by
\begin{align}\label{f0=0}
        \epsilon(\pmb{x})=e^{\frac{1}{2}\omega(\pmb{x})} {\epsilon_{-}(x^{-})  \choose \epsilon_{+}(x^{+})}= {e^{\frac{1}{2}(\omega_{+}(x^{+})-\omega_{-}(x^{-}))} \epsilon_{-}^{(0)}  \choose e^{-\frac{1}{2}(\omega_{+}(x^{+})-\omega_{-}(x^{-}))} \epsilon_{+}^{(0)} }\,,
\end{align}
where $x^{\pm}\equiv t\pm x$, $\omega(\pmb{x})=\omega_{+}(x^{+})+\omega_{-}(x^{-})$ and $\epsilon_{\pm}^{(0)}$ are constant Grassmann variables. See Appendix \ref{AppB} for detailed derivations. From this expression of $\omega(\pmb{x})$, we can read $\mathcal{R}=0$ which means that the background spacetime is  flat. Since the Grassmann variables $\epsilon_{\pm}^{(0)}$ are independent of each other, there are two real supersymmetries\footnote{Strictly speaking, this can be seen as an abuse of terminology. However, considering the similar structure in the solution of $\epsilon$, we intend to refer this situation as ``rigid Rindler supersymmetry'', in line with the terminology of ``rigid AdS$_2$ supersymmetry'' used in \cite{Beccaria:2019dju}.
When we use the term of supercharge, it is well-defined in the context of conventional conserved charge.

} which corresponds to $\mathcal{N}=(1,1)$. As a concrete example, we present the results for Rindler spacetime which are
\begin{align}\label{Rinw(x)}
        \epsilon(\pmb{x})= {e^{\frac{1}{2}bt } \epsilon_{-}^{(0)}  \choose e^{-\frac{1}{2}bt} \epsilon_{+}^{(0)} }\,,\quad \omega(\pmb{x})=bx=\frac{b}{2}x^{+} -\frac{b}{2}x^{-}\,,
\end{align}
with a constant acceleration parameter $b$. The supersymmetric action in  Rindler spacetime is
\begin{align} \label{Rindler}
        S_{\rm Rindler}=\int d^2x \sqrt{-g}\mathcal{L}_0\,,\quad \sqrt{-g}=e^{2bx}\,,
\end{align}
where $\mathcal{L}_0$ was given in \eqref{L0} with replacement of $W(\phi)$ by $\mathcal{W}(\phi,\mathcal{R})$.
The supersymmetrization of a free theory in Rindler spacetime was shown using the (quantum) Hamiltonian formalism~\cite{vanHolten:2020rsb}. On the other hand, in our approach, not only a free theory but also the $\mathcal{N}=(1,1)$ Wess-Zumino model can be constructed in Rindler spacetime.
In the next section, we will revisit the spacetime-dependent supersymmetry variation parameter from the IFT viewpoint.

Next, we consider the case of nonzero constant $f_0$, $f_0\neq 0$. In this case, the solution to the Killing spinor equation \eqref{mainKS} is given by
\begin{align}\label{f0ne0}
        \epsilon(\pmb{x})=e^{\frac{1}{2}\omega(\pmb{x})}{\epsilon_{(1)}+x^{-}\epsilon_{(2)}  \choose \epsilon_{(1)}+x^{+}\epsilon_{(2)}}\,,
\end{align}
where $\epsilon_{(1)}$ and $\epsilon_{(2)}$ are constant Grassmann variables. Here, $\omega(\pmb{x})$ should satisfy
\begin{align}
        2x\partial_{\mp}\omega=\pm 1\,,\quad \partial_{+}\partial_{-}\omega= \frac{1}{2}f_0e^{\omega}\partial_{-}\omega
\,=\frac{1}{2}f_0e^{\omega}\partial_{+}\omega
\,.
\end{align}
Solving above equations, we obtain
\begin{align}\label{AdS}
        e^{\omega (x)} = \frac{\ell}{x}\,, \qquad f_0=-\frac{1}{\ell}\,,
\end{align}
with a dimensionful parameter $\ell$, $\ell>0$. In this case, the metric represents nothing but a Poincar\'{e} patch of AdS$_{2}$, $\ell$ denotes the AdS radius, and $\mathcal{R}=-\frac{2}{\ell^2}$. Since the Grassmann variables $\epsilon_{(1)}$ and $\epsilon_{(2)}$ are independent of each other, there are two real supersymmetries, {\it i.e.}, $\mathcal{N}=(1,1)$ \cite{Beccaria:2019dju}.
So the supersymmetric scalar field action in the AdS background, which reads from \eqref{L2}, is given by 
\begin{align}\label{SAdS}
S_{{\rm AdS}} =\int d^2 x \sqrt{-g} \left( \mathcal{L}_0 + \frac{1}{\ell} W(\phi)\right),
\end{align}
where  $W(\phi) = \mathcal{W} (\phi, \mathcal{R} = -\frac{2}{\ell^2})$.
In this way, we have reproduced the supersymmetric action \eqref{SAdS} in the AdS$_2$ background which was given in \cite{Higashijima:1983wy, Bardeen:1984hm, Beccaria:2019dju}. Since \eqref{AdS} is a unique solution for $f_0={\it constant}\neq0$, we check that de Sitter spacetime cannot be a supersymmetric background.  It is obvious from the relation \eqref{mainKS2} that if  $\mathcal{R}$ is a constant, $\mathcal{R}$ should be  non-positive. 
\subsection{$f(\mathcal{R})\neq$ {\it constant}} \label{222}

In this case, the solution to the Killing spinor equation \eqref{mainKS} is given by
\begin{align}\label{epsilon}
        \epsilon(\pmb{x})= e^{\frac{1}{2}\omega(\pmb{x})} {p_{-}(x^{-})\epsilon_{0}  \choose p_{+}(x^{+})\epsilon_{0} }\,,
\end{align}
where $\epsilon_{0}$ is a constant Grassmann variable and $p_{\pm}(x^{\pm})$ depend only on a single null coordinate, respectively.

The Killing spinor equation \eqref{mainKS} gives us
\begin{align}
        -\frac{1}{2}(\mathcal{R}+2f^2)\epsilon=\nabla_{\mu}f\gamma^{\mu}\epsilon=\frac{\partial f}{\partial \mathcal{R}}\nabla_{\mu}\mathcal{R}\gamma^{\mu}\epsilon \,,
\end{align}
which restricts the Ricci scalar as $\partial_{+}\mathcal{R} \partial_{-} \mathcal{R} <0$.
  This condition tells us that the solution to the Killing spinor equation does not exist, when $\mathcal{R}$ depends only on the time coordinate, $t$. 
  In fact, by constructing Killing vector as forming bilinears out of classical commuting Killing spinors, one can see the existence of time-like Killing vector.
    Then we focus on the case that $\omega$ depends only on the spatial coordinate, $x$. One can show that $p_{+}^2=p_{-}^2=$1 after rescaling $\epsilon_0$ and
\begin{align}\label{sols}
        \epsilon(x) = e^{\frac{1}{2}\omega(x)} {\,\epsilon_0 \choose \pm\epsilon_0} ,\qquad f(\mathcal{R}) =\pm \omega'  e^{- \omega}\,,
\end{align}
where ${}'\equiv\frac{\partial}{\partial x}$ . 
 See the Appendix \ref{AppB} for detailed derivation. In the following, we focus on the upper sign in the above relations without loss of generality.

Due to the projection condition $\gamma^{\hat{x}} \epsilon(x) = \epsilon(x) $  for the supersymmetry variation parameter, the number of real supersymmetries reduces to one. This projection condition is identical with that of the supersymmetric IFT~\cite{Kwon:2021flc}. Therefore, we can say that the supersymmetry  in this case is $\mathcal{N} = \big( \frac{1}{2}, \frac{1}{2}\big)$.


\subsection{A  supersymmetric background} \label{24}

In this subsection, we  present a concrete example for $f(\mathcal{R})\neq{\it constant}$  with $\mathcal{R} = \mathcal{R}(x)$. Specifically, we take \eqref{sols} as
\begin{align}\label{fR}
        f(\mathcal{R})=\frac{\xi}{m_0} \mathcal{R}\,,
\end{align}
where $\xi$ is a dimensionless parameter and $m_0$ is introduced for a dimensional reason. The differential equation \eqref{sols} for $\omega(x)$ with the upper sign becomes
\begin{align}\label{diffw}
  \omega''   + \frac{m_0}{2 \xi}\, e^{ \omega} \omega'=0\,.
\end{align} 
A solution to the differential equation is given by 
\begin{align}
e^{\omega(x)}=\frac{1}{ a+ e^{-bx}}\,, \label{susybackgd}
\end{align}
where $a$ and $b$ are constants satisfying $ab = \frac{m_0}{2\xi}$. Here, $m_0$ and $\xi$ are independent Lagrangian parameters, while $a$ and $b$ are solution parameters which are not independent of each other. Therefore, the number of free parameters in the solution \eqref{susybackgd} is one. See Appendix \ref{AppC} for a possible Lagrangian allowing this solution.
Without loss of generality, we can take $b>0$ by using the reflection symmetry of  $x$-coordinate of our background spacetime. Since the corresponding curvature scalar is given by 
\begin{align}
\mathcal{R} = 2 ab^2 e^{-bx},
\end{align}
the geometry has a naked null curvature singularity at $x \to -\infty $ for $a\neq0$. As shown in Appendix \ref{AppGeodesics}, the spacetime described by \eqref{susybackgd} is geodesically incomplete.    However, this singularity  is a {\it mild}  one as will be shown in Section \ref{wavedy}.

\begin{figure}[http]
   \centering
 \subfigure[]{    \includegraphics[scale=0.8]{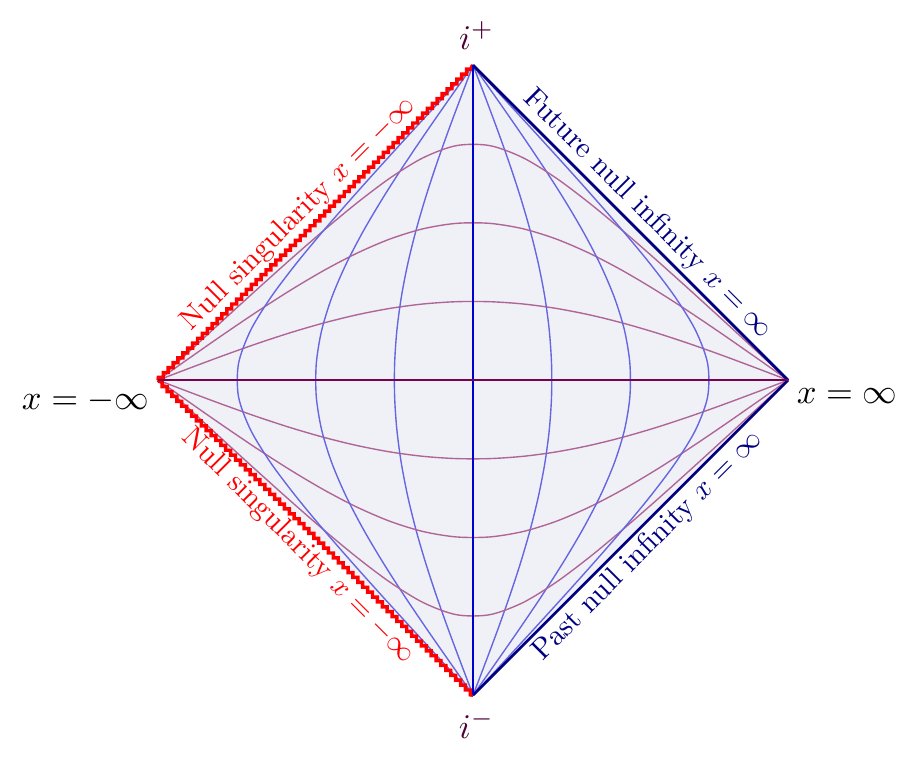} }
 \hfill
   \subfigure[]{    \includegraphics[scale=0.8]{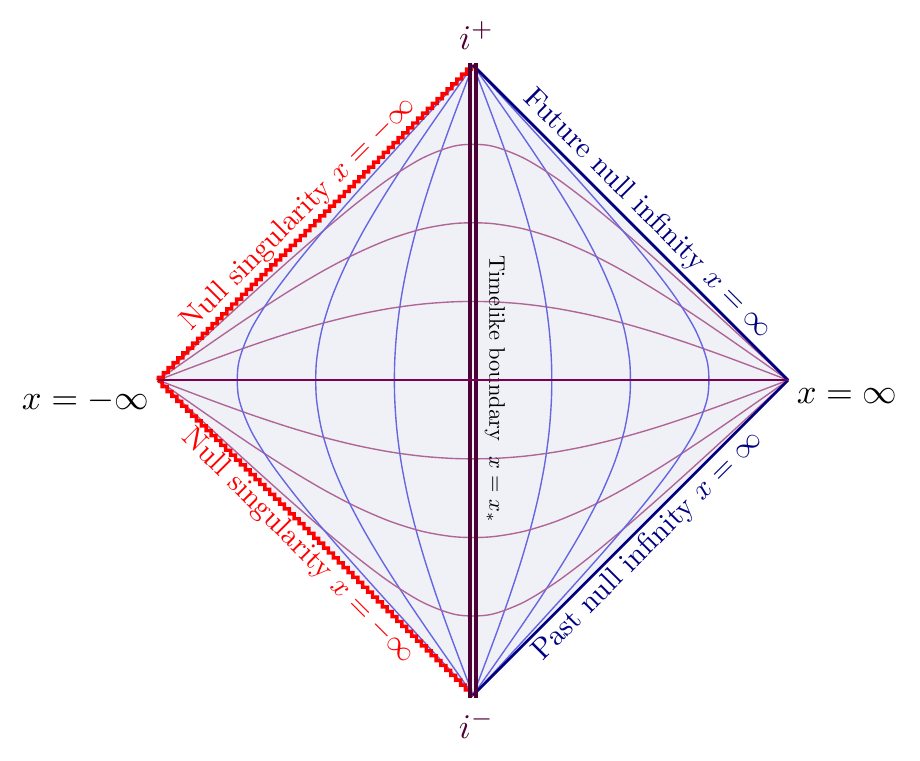} }  
\caption{$(a)$ Penrose diagram for \eqref{susybackgd}  with $a>0$.  $(b)$ Penrose diagram for \eqref{susybackgd}  with $a<0$. The red zigzag line denotes naked null singularity. The blue thick line does null infinity. The causally disconnected $L$ and $R$ parts in the Fig.~\ref{PenroseD}($b$) are drawn in one diagram. One should keep in mind that  the double solid line denotes the boundary of two parts, which is not traversable.}
\label{PenroseD}
\end{figure}

In the case of $a=0$, the spacetime becomes flat, which is nothing but Rindler spacetime in \eqref{Rindler}. In the case of $a<0$, the spacetime is split into causally disconnected two parts, $L$ and $R$, along the hypersurface $x=x^*\equiv -\frac{1}{b} \ln|a|$; The $L$ part includes the singularity while the $R$ part does the asymptotically flat region. See the Penrose diagram in Fig.~\ref{PenroseD}($b$). On the other hand, the spacetime with $a>0$ has only one causally connected region including a naked null singularity. 
See Fig.~\ref{PenroseD}($a$).
In Section \ref{SIFT}, we present an IFT interpretation of these geometries, and in Section \ref{wavedy}, we analyze wave dynamics on the geometry of \eqref{susybackgd}.

Some comments are in order.
It is interesting to note that a metric with naked null singularity 
\begin{align}
ds^2 = - \left( 1+ \frac{M}{r} \right)^{-2} dt^2 + \left( 1+ \frac{M}{r} \right)^{2} dr^2  + r^2 d \Omega ^2  ~\label{nullksingul}
\end{align} 
has been studied for the purpose to give a description of the shadow effect of the Galactic center M87~\cite{EventHorizonTelescope:2019dse} without an event horizon, in the absence of a photon sphere and a thin shell matter \cite{Joshi:2020tlq}. 
The metric \eqref{nullksingul} is not a solution to the Einstein equation with a minimally coupled
scalar field, but it has been shown that timelike and null-like naked sigularities can be formed as the final state
of gravitational collapse of an inhomogeneous matter cloud~\cite{Joshi:1993zg,Hyun:2006xt}.
When $S^2$ part is ignored, our background spacetime metric \eqref{susybackgd} for $a>0$ has the same global causal structure\footnote{In \eqref{nullksingul}, the naked null singularity appears at $r \rightarrow 0$, while  the metric becomes asymptotically flat as $r\rightarrow\infty$. These asymptotic behaviors are same with those of the metric \eqref{susybackgd} with $a>0$, which may be understood by setting $r\sim e^{x}$.} with the above metric and a naked null singularity. In this sense, it would be interesting to study the shadow effect in our supersymmetric background with expectation of advantages of solvability.

\subsection{Various field theories on the supersymmetric background}\label{25}

We consider the background depending only on the spatial coordinate, $x$, {\it i.e.} $\mathcal{R}=\mathcal{R}(x)$. In this case, we can rederive all the relevant equations obtained from the Killing spinor equations by deriving the energy bound.

To derive this bound, firstly, we write the bosonic part of the canonical Hamiltonian as
\begin{align}\label{Hcan}
        \sqrt{-g}\mathcal{H}_{\rm can}&=\frac{1}{2}\left[ \dot{\phi}^2 + \phi'{}^2 +e^{2\omega}\left\{ \left(\frac{\partial \mathcal{W}}{\partial \phi}\right)^2 +2f(\mathcal{R})\mathcal{W} (\phi,\mathcal{R})+2\,\mathcal{U}(\phi,\mathcal{R}) \right\} \right]\nonumber \\
        &= \frac{1}{2}\left[\dot{\phi} ^2 +\left(\phi'\mp e^{\omega}\frac{\partial \mathcal{W}}{\partial \phi} \right)^2  \pm 2e^{\omega}\phi'\frac{\partial \mathcal{W}}{\partial\phi} +2e^{2\omega}\left\{f(\mathcal{R})\mathcal{W}(\phi,\mathcal{R}) +\,\mathcal{U}(\phi,\mathcal{R}) \right\}\right].
\end{align}
We can make the last three terms in the second equality in \eqref{Hcan} to be a total derivative form by requiring conditions for $f$ and $\mathcal{U}$,
\begin{align}\label{cond4}
        f(\mathcal{R})=\pm\omega'e^{-\omega}\,,\qquad \mathcal{U}=\pm e^{-\omega} \mathcal{R}'\frac{\partial \mathcal{W}}{\partial \mathcal{R}} \,.
\end{align}
Then, the  energy is bounded from below as
\begin{align}\label{Ebound}
        \sqrt{-g}\mathcal{H}_{\rm can}\geq\pm \left(e^{\omega} \mathcal{W} \right)'\,.
\end{align}
The conditions  in \eqref{cond4}   are identical with those  from Killing spinor equations. The first condition in \eqref{cond4} is identical with  \eqref{sols}.   Using $\mathcal{R}=-2\omega''e^{-2\omega}$ and the first condition in \eqref{cond4}, one can write the second condition in \eqref{cond4}  in the form of 
$$\,\mathcal{U}=-\frac{1}{2}(\mathcal{R}+2f^2)\frac{\frac{\partial\mathcal{W}}{\partial\mathcal{R}} }{\frac{\partial f}{\partial\mathcal{R}}}\,,$$
which is the same expression from  \eqref{U} by applying \eqref{G_n}.
We note that the above procedure to make a total derivative in the canonical Hamiltonian gives the constraint equations for supersymmetric backgrounds.

 As in the spinor case, we choose the upper sign in \eqref{cond4} without loss of generality.
To saturate the equality with the upper sign in \eqref{Ebound}, we should require  the completed square terms in \eqref{Hcan}   to vanish \begin{align} \label{BPS}
        \dot{\phi}=0\,,\qquad \phi'- e^{\omega}\frac{\partial \mathcal{W}}{\partial\phi} =0\,,
\end{align}
which are the so-called BPS equations.
These equations can also be obtained from the fermion variation
\begin{align}
        \delta \Psi=-\gamma^{\mu}\nabla_{\mu}\phi\, \epsilon +\frac{\partial \mathcal{W}}{\partial\phi}\epsilon=0\,.
\end{align}
One can see that even in the curved background the BPS solution is static, as we read from \eqref{BPS}.

As we have discussed previously,  SFTCS in \eqref{L2} encompasses  known supersymmetric field theories in flat and AdS$_2$ spacetimes. In these two backgrounds satisfying $\mathcal{F}_n(\mathcal{R}) = \lambda_n = {\it constant} $,  one can construct various models by choosing appropriate superpotentials $W(\phi)=\mathcal{W}(\phi,\mathcal{R})$, for example,
\begin{align}
W(\phi) = \left\{ \begin{array}{cl}
        \frac{1}{2}m_{0}\phi^2\,,  &\text{free theory}  \\
        \frac{\lambda_{\phi}}{4}(\phi^4-2d\phi^2), &\text{$\phi^6$-theory} \\
        \frac{\lambda_{\rm L}}{2c}e^{2c\phi}\,, &\text{Liouville} \\
        -\frac{\lambda_s}{\beta^2}\cos\frac{\beta\phi}{2}\,, & \text{Sine-Gordon}
\end{array}     \right.
\end{align}
with dimensionful constants $m_0,\,\lambda_s,\,\lambda_{\rm L},\,\lambda_\phi$ and dimensionless constants, $\beta,\,c,\,d.$

%


Just like SFTCSs on the flat and AdS backgrounds which have two supersymmetries, known as $\mathcal{N} = (1,1)$, one can also construct above models in our supersymmetric background \eqref{susybackgd}, which have only one supercharge, referred as $\mathcal{N} = \big(\frac{1}{2}, \frac{1}{2}\big)$. In fact, we can consider more general superpotentials of non-constant arbitrary $\mathcal{F}_{n}(\mathcal{R})$ with which we can construct SFTCSs. In this case, the supersymmetric Lagrangian contains another term $\mathcal{U}(\phi,\mathcal{R})$ given in \eqref{U}.

As a specific SFTCS on the supersymmetric background \eqref{susybackgd} for which $f(\mathcal{R})$ is taken as \eqref{fR},
let us consider a free theory. We mean the free theory to take $\mathcal{F}_{n\neq2}(\mathcal{R})=0$, $\mathcal{F}_{2}(\mathcal{R})=\frac{m_0}{2}$, and then  $\mathcal{U}(\phi,\mathcal{R})=0$.
 In this free case, the bosonic Lagrangian is given by
\begin{align} \label{Lfree}
        \sqrt{-g}\, \mathcal{L}_{\rm free}=\sqrt{-g}\bigg[ -\frac{1}{2}g^{\mu\nu}\nabla_{\mu}\phi\nabla_{\nu}\phi-\frac{1}{2}m_0^2\phi^2 -\frac{1}{2}\xi\mathcal{R}\phi^2\bigg] \,.
\end{align}
Since this free theory is constructed on the curved background, we need to figure out the field configuration saturating the energy bound \eqref{Ebound}. That is to say, we need to investigate BPS equations which are given by
\begin{align}\label{BPSfree}
        \dot{\phi}=0\,,\qquad\phi'-m_0e^{\omega}\phi=0\,.
\end{align}
By solving the above BPS equations in our background, we obtain the BPS solution for the free theory as\footnote{The non-normalizability of zero modes associated with the shift symmetry $\phi \rightarrow \phi + \phi_{\rm BPS}$  in $L^{2}$-norm is a subtle issue in our context. The related topic was previously discussed in~\cite{Griffin:2014bta}, where it was explored in the context of a polynomial function shift symmetry in scalar field theory. Further investigation of the shift symmetry in our model is required to  understand its implications.}
\begin{align}
        \phi_{\rm BPS}(x) = \phi_0 (ae^{bx}+1)^{2\xi}\,,\qquad 2\xi=\frac{m_0}{ab}\,.
\end{align}

We consider two cases separately: $\xi=0$ and $\xi\neq0$. At first, when $\xi=0$, {\it i.e.} $m_0=0$ with $a\neq0$ (Note that we have chosen $b>0$), $\phi_{\rm BPS}(x)$ is a constant as can be seen in \eqref{BPSfree}, and then $E_{\rm BPS}=0$. 
In the next, we consider the case of $\xi\neq0$ according to the sign of $a$. In the case of $a>0$, the energy for the BPS solution is given by
\begin{align}
        E_{\rm BPS}=\int_{-\infty}^{+\infty} dx \sqrt{-g}\, \mathcal{H}_{\rm can}\Big|_{\rm BPS}=\frac{1}{2}m_0\phi_0^2e^{bx}(ae^{bx}+1)^{4\xi-1}\Big|_{-\infty}^{+\infty}\,.
\end{align}
Now, we classify BPS solutions according to the Lagrangian parameter $\xi$.
\begin{enumerate}
        \item $\xi> 0$: Requiring finite energy solution, we should take $\phi_0=0$. In this case, the lowest energy solution is unique.
        \item $\xi<0$: The solution family having the lowest energy is parametrized with one parameter, $\phi_0$.
\end{enumerate}
In both cases, the BPS solutions have zero energy, {\it i.e.} $E_{\rm BPS}=0$.

In the case of $a<0$, the spacetime is composed of two parts, $L$ and $R$ causally disconnected along $x_*$. 
See Fig.~\ref{PenroseD}($b$). Furthermore, the reality condition of $\phi_{\rm BPS}(x)$ gives constraints on $\xi$, because $ae^{bx}+1<0$ in the $R$ part. For instance, if $2\xi$ is taken to be an integer,  the reality of $\phi_{\rm BPS}(x)$ is guaranteed. In general, $\phi_{\rm BPS}(x)$ becomes complex or multi-valued for a generic value of $\xi$, which we will not consider in this paper. Within this category, the BPS energy expressions for each part are given respectively by
\begin{align}
        E_{\rm BPS}^{L}&=\int_{-\infty}^{x_*} dx \sqrt{-g}\, \mathcal{H}_{\rm can}\Big|_{\rm BPS}=\frac{1}{2}m_0\phi_0^2e^{bx}(ae^{bx}+1)^{4\xi-1}\Big|_{-\infty}^{x_*}\,,\\
    E_{\rm BPS}^{R}&=\int_{x_*}^{+\infty} dx \sqrt{-g}\, \mathcal{H}_{\rm can}\Big|_{\rm BPS}=\frac{1}{2}m_0\phi_0^2e^{bx}(ae^{bx}+1)^{4\xi-1}\Big|_{x_*}^{+\infty}\,.
\end{align}
We will now categorize BPS solutions based on the Lagrangian parameter $\xi$.
\begin{enumerate}
        \item $\xi>\frac{1}{4}$: In the $R$ part, requiring a finite energy solution, we should take $\phi_0=0$. The solution of the lowest energy, $E_{\rm BPS}^{R}=0$, is unique. In the $L$ part, the solution family having the lowest energy, $E_{\rm BPS}^{L}=0$, is parametrized with one parameter, $\phi_0$.
        \item $\xi=\frac{1}{4}$: In this case, $m_{0} <0$.   In the $L$ part, there is a solution family labeled by $\phi_{0}$ since  $E^{L}_{\rm BPS} = \frac{1}{2}m_{0} \phi^{2}_{0} e^{bx_{*}}  = - b\xi \phi^{2}_{0}< 0$, while we should take $\phi_{0}=0$ in the $R$ side for a finite $E^{R}_{\rm BPS}$, and then $E^{R}_{\rm BPS}=0$.
        \item $\xi<\frac{1}{4},\,\xi\neq0$: To ensure the finiteness of $E^{L/R}_{\rm BPS}$ in both of $L/R$ parts,  we should take $\phi_0=0$. Here, $E_{\rm BPS}^{L/R}=0$.       
\end{enumerate}

Finally,  in the limit of  $a\rightarrow 0^{+}$, {\it i.e.} Rindler case, together with $m_{0}\rightarrow 0^{+}$,  the BPS energy expression is
\begin{align}
        E_{\rm BPS}=\int_{-\infty}^{+\infty} dx \sqrt{-g}\, \mathcal{H}_{\rm can}\Big|_{\rm BPS} \longrightarrow 0 \,.
\end{align}
Here, there is a family of classical solutions parametrized by $\phi_{0}$ as before. 

One may notice that $E_{\rm BPS}=0$ holds for all $\xi$ values except $\xi=\frac{1}{4}$, and is independent of the sign of $a$. When $\xi=\frac{1}{4}$, $E_{\rm BPS}^{L}$ becomes negative. However, this does not pose a problem in the context of supersymmetric field theory with only one supercharge \cite{Kwon:2021flc}.

We explore time-dependent physics of the bosonic part of this free theory in the supersymmetric background \eqref{susybackgd} in Sec.~\ref{wavedy}. In that section, we explain the meaning of the {\it mild} singularity in the view point of scalar wave dynamics. In Sec. \ref{wavedy}, we focus on the case of $a>0$ and $\xi \ge \frac{1}{4}$.


\section{Supersymmetric Inhomogeneous Field Theory}\label{SIFT}

As  explained in the previous section, we have extended  the relation between FTCSs and corresponding IFTs in $(1+1)$ dimensions proposed in~\cite{Ho:2022ibc} to the supersymmetric 
setup and obtained a specific supersymmetric background  on which  various field theories can live. 
In this section, we interpret the results 
of SFTCS
on that background
in the language of $(1+1)$-dimensional SIFT with spacetime-dependent couplings, which lives in  Minkowski background.
Then, we focus on a free SIFT as a specific example in which
 the supersymmetric quantum mechanics can be embedded.

\subsection{Model construction}

In the previous section, we show that SFTCS can have two real supersymmetries in the case of $f(\mathcal{R})=constant$.
Motivated by the bosonic relation between  FTCS and IFT in $(1+1)$ dimensions,  SIFT  with two real supersymmetries corresponding to  SFTCS can also be taken into account.
Indeed, we show how to construct the SIFT models with two real supersymmetries which can be thought as an extension of  
 the previous SIFT construction  with one real supercharge in~\cite{Kwon:2021flc}.

In  Minkowski background,  conventional  $\mathcal{N}=(1,1)$ Wess-Zumino models \eqref{SFT1} with two real supersymmetries have constant coupling parameters $\lambda_n$'s. On the other hand,  position-dependent coupling parameters lead to  $\mathcal{N}= (\frac{1}{2}, \frac{1}{2})$  SIFTs  with one real supercharge. To cover the case of two real supersymmetries in SIFT,
we need to consider spacetime-dependent  couplings rather than position-dependent ones. That is to say, the Lagrangian \eqref{SFT1} is to be  generalized  with the extended superpotential $W(\phi, \pmb{x})$ as
\begin{align}\label{W(phix)}
W(\phi) = \sum_{n\ge 2}\lambda_n \phi^n \,\, \longrightarrow\,\, W(\phi, \pmb{x}) = \sum_{n\ge 2}\lambda_n (\pmb{x})\phi^n ,
\end{align}
where $\pmb{x} = (t,x)$.  As a next step,  performing supersymmetric variation given by 
\begin{align}\label{svar2}
        \delta \phi &= i\bar{\psi}\varepsilon\,, \nonumber \\
        \delta \psi &= -\gamma^{\mu}_{{\rm F}}\partial_{\mu}\phi~ \varepsilon + \Big( \frac{\partial W(\phi, \pmb{x}) }{\partial \phi}\Big) \varepsilon\,,
\end{align}
we obtain 
\begin{align}\label{dLSFT}
\delta \mathcal{L}_{{\rm SFT}}|_{W(\phi) \to W(\phi, \pmb{x})} &= - i \partial_{\mu} \phi \bar \psi \big( \eta^{\mu\nu} 
- \gamma^{\mu \nu}_{\rm F} \big) \partial_{\nu} \varepsilon
+  i \bar{\psi} \gamma^{\mu}_{\rm F}  (\partial_{\mu}\varepsilon)\frac{\partial W}{ \partial \phi} + i \bar{\psi} \gamma^{\mu}_{\rm F} \varepsilon \partial_{\mu} \Big(\, \frac{\partial W}{ \partial \phi} \Big),
\end{align}
where $ \partial_{\mu} \big(\, \frac{\partial W}{ \partial \phi} \big) \equiv \sum_{n\ge 2} n (\partial_{\mu} \lambda_n) \phi^{n-1}$ and the supersymmetric parameter $\varepsilon$ depends on spacetime coordinates, $\varepsilon = \varepsilon(\pmb{x})$.

As we see in \eqref{dLSFT},  the Lagrangian $\mathcal{L}_{{\rm SFT}}|_{W(\phi) \to W(\phi, \pmb{x})}$ is not invariant under the supersymmetric  variation in \eqref{svar2}.  To make supersymmetric completion,\!\!\footnote{Our construction of SIFT with {\it spacetime-dependent} couplings is  a generalization of the supersymmetrization procedure in \cite{Kwon:2021flc}. } we consider the most general form of Killing spinor equation in flat spacetime as
\begin{align}\label{genKS2}
\partial_{\mu} \varepsilon = h_{\mu} \varepsilon + h_{\mu\nu} \gamma^{\nu}_{\rm F} \varepsilon + h_{\mu\nu\rho} \gamma^{\nu\rho}_{\rm F}\varepsilon,
\end{align}
where $h_{\mu}$, $h_{\mu\nu}$, and $h_{\mu\nu\rho} = -h_{\mu\rho\nu}$ are spacetime-dependent functions. Note that the Killing spinor equation reduces to
\begin{align}\label{KSIFT}
\partial_{\mu} \varepsilon = \frac{1}{2} h \gamma_{\mu}^{\rm F} \varepsilon+ h_\mu  \,\varepsilon+h_\nu  \,\gamma^{\rm F}_{\;\mu}{}^{\nu}\varepsilon\,,\qquad h\equiv (\eta^{\mu\nu}-\gamma_{\rm F}^{\mu\nu})h_{\mu\nu}\,.
\end{align}
The unique feature of this Killing spinor equation in SIFT is that $h$-tensors do not vanish in general.
All the $h$-tensors are zeros in usual supersymmetric field theories on flat spacetime.   
To achieve the supersymmetric completion from \eqref{dLSFT}, an additional condition should be satisfied as
\begin{align}\label{g_n}
\gamma^{\mu}_{\rm F}  \big(\partial_{\mu} \lambda_n (\pmb{x})+2h_{\mu}(\pmb{x})\lambda_{n}(\pmb{x})\big) \varepsilon (\pmb{x}) = g_n(\pmb{x}) \varepsilon (\pmb{x}). 
\end{align}
Then, we formally obtain a supersymmetric invariant Lagrangian
\begin{align}\label{LSIFT}
\mathcal{L}_{{\rm SIFT}} = \mathcal{L}_{{\rm SFT}}|_{W(\phi) \to W(\phi, \pmb{x})}-  \sum_{n\ge 2} \big( h\lambda_n  + g_n  \big)\phi^n  
\end{align}
under the supersymmetric variation \eqref{svar2}. 
In the following, for simplicity, we will take $h_{\mu}=\partial_{\mu}q $  with a spacetime-dependent function $q=q(\pmb{x})$. 

Now, we try to find several explicit examples of SIFT constructed in the above. After solving the Killing spinor equation \eqref{KSIFT}, we observe that there are three classes of solutions categorized by the form of $e^{2q}h$. Firstly, in the case of $h=0$,  the solution is given by 
\begin{align}\label{flate}
\varepsilon= {e^{2q_{-}(x^-)} \varepsilon_{-}^{(0)}  \choose e^{2q_{+}(x^+)} \varepsilon_{+}^{(0)} }\,.
\end{align}
Secondly, in the case $e^{2q}h=constant\neq0$,  it is given by
\begin{align}
\varepsilon={\varepsilon_{(1)}+x^{-}\varepsilon_{(2)}  \choose \varepsilon_{(1)}+x^{+}\varepsilon_{(2)}}\,.
\end{align}
Finally, in the other case of $e^{2q}h(\neq constant)$, it is
\begin{align}
\varepsilon= {p_{-}(x^{-})\varepsilon_{0}  \choose p_{+}(x^{+})\varepsilon_{0} }\,.
\end{align}
In the above solutions, $\varepsilon_{\pm}^{(0)},\,\varepsilon_{(1,2)},$ and $\varepsilon_{0}$ are constant Grassmann variables. 
Here, $q_{\pm}$ and $p_{\pm}$ are real valued functions depending on a single null coordinate, $x^{\pm}$, respectively. In the first case, $q(\pmb{x})$  is decomposed into two functions, $q_+$ and $q_-$ whose  are functions of $x^+$ and $x^-$, respectively,  $i.e.$  $q(\pmb{x})=q_{+}(x^+)+q_{-}(x^-)$. In the final case, the condition satisfied by $p_{\pm}(x^{\pm})$ is given in Appendix \ref{AppB}.
 As can be seen from the independent number of Grassmann variables, the first and the second solutions correspond to two real supersymmetries, while the last case does to one real
charge. 

Solving \eqref{g_n}, we have the relation
\begin{align}
g_n^2(\pmb{x}) = - (\dot{\lambda}_n+2\dot{q}\lambda_n)^2 +(\lambda_n{}'+2q'\lambda_n)^2 .
\end{align}
To have real value of $g_n$, the inhomogeneous coupling function $\lambda_n$ should satisfy the condition $ - (\dot{\lambda}_n+2\dot{q}\lambda_n)^2 +(\lambda_n{}'+2q'\lambda_n)^2  \ge 0$. When $\lambda_n$ and $q$ depend only on time, there is not any supersymmetric solution. 
To avoid possible complication, we consider only position-dependent couplings. In this case, we can set $p^2_{\pm}=1$. Then, we obtain
\begin{align}
h(\pmb{x})=\mp2q'(\pmb{x})\,,
\end{align}
where the $\mp$ sign comes from the projection condition, $\gamma_{\rm F}^{x}\varepsilon=\pm\varepsilon$. In the following, we take the upper sign without loss of generality. 

Now, we are in position of summarizing what is extended from the previous work~\cite{Kwon:2021flc}, where the coupling parameters depend only on the spatial coordinate and the Killing spinor equation is $\partial_{\mu}\varepsilon=0$. Here, we  introduce additional functions, $h$ and $h_\mu$ in the right-hand side of \eqref{KSIFT}, so that we unveil models of two real supersymmetries, as well as those of one real supercharge with spacetime-dependent parameters in SIFT.
When coupling parameters depend only on the position, the Lagrangian with one real supercharge was shown to be
\begin{align}\label{LagOP}
        \mathcal{L}_{\rm SIFT}=\mathcal{L}_{\rm SFT}\Big|_{\lambda_n \to \lambda_n(x)}-\frac{\partial W(\phi,x)}{\partial x}\,,
\end{align}
which is the special case of the Lagrangian \eqref{LSIFT} with $g_n(x)=\lambda_n'(x)$ and $h=q=0$. 

For models of $h=0$ and $q\ne 0$  with two real supersymmetries, the Lagrangian parameters $\lambda_{n}$ and $g_n$ are given respectively by
\begin{align}
        \lambda_{n}(\pmb{x})= \lambda_n^{(0)} e^{-2q(\pmb{x}) },  \qquad q(\pmb{x})=q_{+}(x^+)+q_{-}(x^-), \qquad g_{n}(\pmb{x})=0\,.
\end{align}
Then, the Lagrangian becomes 
\begin{align}
\mathcal{L}_{\rm SIFT}=\mathcal{L}_{\rm SFT}\Big|_{\lambda_n \to\lambda_n^{(0)} e^{-2q(\pmb{x})}} . \end{align}
As a special example, let us consider SIFT corresponding to SFTCS on Rindler spacetime, where $q_{\pm} =- \frac{b}{4}\, x^{\pm}$. In this case, the superpotential is given by  
\begin{align}
W(\phi, \pmb{x}) = e^{bx} \sum_{n\ge 2} \lambda_n^{(0)} \phi^n =e^{\omega(x)} \mathcal{W}(\phi, \mathcal{R} =0),
\end{align}
where $\omega(x) = bx$ in Rindler spacetime  \eqref{Rinw(x)}.
In spite that all coupling parameters depend only on position $x$, the SIFT Lagrangian does not belong to the category of \eqref{LagOP}.  However, if we relax our choice of $f=f(\mathcal{R})$, we cannot exclude the possibility that it may fall within that category of \eqref{LagOP}.
On the other hand, in the model of $e^{2q}h=-C=constant\neq0$ with two real supercharges, we obtain
\begin{align}\label{lamn}
        \lambda_{n}(\pmb{x})= \frac{c_n}{x} \,, \qquad g_n(\pmb{x})=0\,,
\end{align}
where $e^{2q} = \frac{x}{C}$ and $c_n$'s are constants, the Lagrangian is given by
\begin{align}
        \mathcal{L}_{\rm SIFT}=\mathcal{L}_{\rm SFT}\Big|_{\lambda_n \to \frac{c_n}{x}}+\frac{1}{x^2}\sum_{n\geq2}c_n\phi^n=\mathcal{L}_{\rm SFT}\Big|_{\lambda_n \to \frac{c_n}{x}}-\frac{\partial W(\phi,x)}{\partial x}\,.
\end{align}
If $C$ is positive(negative), the range of coordinate $x$ is restricted to positive(negative). This  SIFT model corresponds to SFTCS on  AdS$_2$ spacetime. As can be seen in the above expression, this model belongs to the same category of \eqref{LagOP}, while supersymmetry enhancement  occurs in the special case of $\lambda_n(x)\propto\frac{1}{x}$.

\subsection{SFTCS and SIFT}

In the previous subsection, we have extended the construction of SIFT in \cite{Kwon:2021flc}. Here, we present the concrete relation between SFTCS and SIFT. 
This is a specific adaptation of the result of \cite{Ho:2022ibc}, in the sense that the correspondence between FTCS in \eqref{FTCSact1} and IFT in \eqref{IFTact1} can be extended to supersymmetric setup within the allowed supersymmetric backgrounds.

At first, we start from the Lagrangian \eqref{L2} of the SFTCS and transform to the Lagrangian \eqref{LSIFT} of the SIFT. Since the dictionary of the transformation for bosonic terms was already constructed in \cite{Ho:2022ibc}, we focus on the transformation of fermionic terms. For a generic metric \eqref{congau} in $(1+1)$ dimensions, the kinetic term for the fermion field $\Psi$ is rewritten as 
\begin{align}
\frac{i}{2} \sqrt{-g} \Psi \gamma^\mu \nabla_{\mu} \Psi = \frac{i}{2}  \bar\psi \gamma^{\mu}_{\rm F} \partial_{\mu} \psi  
\end{align}
under the field redefinition 
\begin{align}\label{Psi=psi}
\Psi = e^{-\frac{\omega}{2}} \psi\,.
\end{align} 
Note that numerical values of flat gamma matrices $\gamma^{t}_{\rm F}$ and $\gamma^{x}_{\rm F}$ are identical with those of tangent space gamma matrices $\gamma^{\hat{t}}$ and $\gamma^{\hat{x}}$, respectively.
By applying the same field redefinition of the fermion field to the fermionic superpotential term and requiring its form invariance as
\begin{align}
\frac{i}{2}\left( \frac{\partial^2 \mathcal{W}}{ \partial \phi^2} \right)\bar{\Psi}\Psi =\frac{i}{2}\left( \frac{\partial^2W}{ \partial \phi^2} \right)\bar{\psi}\psi\,,
\end{align}
we obtain the relation the superpotential $W(\phi, \pmb{x})$ in the SIFT and $\mathcal{W}(\phi,\mathcal{R})$ in the SFTCS as
\begin{align}\label{WfcW}
W(\phi, \pmb{x} ) = e^{w(\pmb{x})} \mathcal{W} (\phi, \mathcal{R}).
\end{align}
Comparing the SFTCS in \eqref{L2} and the SIFT in \eqref{LSIFT}, one can identify relations among couplings in both sides as
\begin{align}\label{ln(x)}
 \mathcal{F}_n (\mathcal{R})= e^{2q(\pmb{x})}\lambda_n (\pmb{x})\,,\quad  f(\mathcal{R})=e^{2q(\pmb{x})}h(\pmb{x})\,,\quad   \mathcal{G}_{n}(\mathcal{R})=e^{4q(\pmb{x})}g_n (\pmb{x})\,,\quad \omega(\pmb{x})=-2q(\pmb{x})\,.
\end{align} 
Note also that the supersymmetry variation parameters in both sides are related as 
\begin{equation} \label{}
\epsilon(\pmb{x}) = e^{-q(\pmb{x})} \varepsilon(\pmb{x})\,.
\end{equation}

Consequently, the SFTCS Lagrangian is converted to the SIFT Lagrangian under the fermion field redefinition \eqref{Psi=psi} and relations of the spacetime dependent coupling in \eqref{ln(x)}. We conclude that our proposal in the previous work~\cite{Ho:2022ibc} can be extended to supersymmetric setup within the allowed supersymmetric backgrounds, {\it e.g.} flat, AdS$_2$, and the spacetime with the metric given in \eqref{susybackgd}.

\subsection{SFTCS/SIFT and SQM}

In the previous subsection, we have shown that there exists a one-to-one correspondence between SFTCS and SIFT in some specific models. Thus, we phrase the following argument in the language of SIFT.
We focus on a free theory with the superpotential given by
\begin{align}
        W(\phi,\pmb{x})=\frac{1}{2}m(\pmb{x})\phi^2\,.
\end{align}

Now, we display several free SIFT Lagrangians according to the value of $e^{2q}h$ in order to show the relation between SIFT and SQM.
When $h=0$, we have shown $m(\pmb{x})=m_0 e^{-2q (\pmb{x})}   $ in the previous section, and so
free SIFT Lagrangian is given by
\begin{align}\label{LSIFT1}
\mathcal{L}_{{\rm SIFT}} =   -\frac{1}{2} \eta^{\mu\nu}\partial_{\mu} \phi \partial_{\nu} \phi + \frac{i}{2} \bar \psi \gamma^{\mu}_{\rm F} \partial_{\mu} \psi + \frac{i}{2} m_0    e^{-2q (\pmb{x})} \psi \psi - \frac{1}{2} m_0^2 e^{-4q (\pmb{x})}\phi^2 \,,
\end{align}
where $q (\pmb{x}) = q_+(x^{+}) + q_-(x^{-})$.
In the case of $e^{2q}h=-C=constant\neq0$, we take $c_2 = \tilde \ell$ and $c_{n\ne 2} = 0$ in \eqref{lamn} such that $m(\pmb{x})=-\frac{\tilde{\ell}}{x}$,  and then  it is given by
\begin{align}\label{LSIFT2}
        \mathcal{L}_{{\rm SIFT}} =   -\frac{1}{2} \eta^{\mu\nu}\partial_{\mu} \phi \partial_{\nu} \phi + \frac{i}{2} \bar \psi \gamma^{\mu}_{\rm F} \partial_{\mu} \psi - i\frac{\tilde{\ell}}{2x}  \bar \psi \psi -  \frac{\tilde{\ell}(\tilde{\ell}+1)}{2x^2}\phi^2 \,.
\end{align}
For a single real supercharge model, we consider the position dependent case only, {\it i.e.} $m(\pmb{x})=m(x)$. Then, the free Lagrangian is
\begin{align}\label{LSIFT3}
        \mathcal{L}_{\rm SIFT}=-\frac{1}{2} \eta^{\mu\nu}\partial_{\mu} \phi \partial_{\nu} \phi + \frac{i}{2} \bar \psi \gamma^{\mu}_{\rm F} \partial_{\mu} \psi + \frac{i}{2} m(x)  \bar \psi \psi - \frac{1}{2} \big(m^2(x)+m'(x)\big)\phi^2 \,.
\end{align}
Note that as we alluded earlier, the supersymmetric Lagrangian \eqref{LSIFT3} includes \eqref{LSIFT2} as a special case with supersymmetry enhancement.

Any SQM reviewed in \cite{Cooper:1994eh} can be embedded in the bosonic part of the above SIFT Lagrangian \eqref{LSIFT3}, which has time translation symmetry. To show this, let us look at the equation of motion of the scalar field mode with a frequency $\omega$,
\begin{equation} \label{eomscalar}
\bigg[ - \frac{d^{2} }{dx^{2}}  +  V_{\rm eff}(x) \bigg] \phi_{\omega}(x) =  \omega^{2}  \phi_{\omega}(x) \,,  \qquad   V_{\rm eff}(x) \equiv m^2(x)+m'(x)\,,
\end{equation}
which is nothing but the Schr\"{o}dinger equation. The interpretation of scalar field equation on the static background in terms of quantum mechanics is a well-known story, but in our case the potential form is exactly that of SQM by taking $m(x)=-W_{\rm QM}(x)$. Here, $W_{\rm QM}$ is the superpotential in SQM~\cite{Cooper:1994eh}. In the field theory viewpoint, one may interpret $V_{\rm eff}$ as an effective mass squared, which can have a negative value. This is natural from the SQM viewpoint and it could lead to interesting consequences\footnote{See the Section~\ref{42}.}.

The free SFTCS on the supersymmetric  background~\eqref{susybackgd} in Section~\ref{SFTCS} can be transcribed as the SIFT in \eqref{eomscalar}. Then, we adopt the technique developed for SQM to investigate scalar wave propagation in that background. 
In the next section, we explore character of naked null singularity of that background through scalar wave dynamics. The so-called shape invariance in SQM plays an interesting role in this regard.

\section{Scalar Wave Dynamics} \label{wavedy}
In this section, we explore a scalar wave propagation on the supersymmetric background \eqref{susybackgd} which has a curvature singularity and is geodesically incomplete. 
See Appendix \ref{AppGeodesics}.  Though there is a curvature singularity  in this supersymmetric background spacetime, it turns out that predictability in  scalar wave  dynamics is not lost, {\it i.e.} the spacetime is globally hyperbolic. In other words, the singularity is a {\it mild} one.  In fact, it has been pointed out that $(1+1)$-dimensional spacetime with a similar null singularity is  globally hyperbolic  \cite{Horowitz:1995gi}. Here, we will show that our singularity doesn't cause serious problems in  scalar wave propagation following the methodology developed in~\cite{Wald:1980jn, Ishibashi:1999vw,Ishibashi:2003jd,Ishibashi:2004wx}. 
From the IFT perspective, this exploration describes equivalently  scalar wave propagation on the inhomogeneous background according to the 
relation~\cite{Ho:2022ibc}.

\subsection{Self-adjoint extension}

%


As stated before, the null singularity in our background metric does not lead to any essential difficulty in scalar wave dynamics. In the following, we provide more details about this singularity issue to clarify this statement. In fact, from the IFT viewpoint there could not exist such a singular behavior of scalar wave propagation. The analysis below about scalar wave dynamics on our supersymmetric background with singularity would support our claim on the relation in supersymmetric setup between SFTCS and SIFT. 

A hypersurface-orthogonal timelike Killing vector is taken as $K_{T}= \frac{\partial}{\partial t}=t^{\mu}\partial_{\mu}$ for  static background. Then, the inner product of two scalar fields on the space-like hypersurface $\Sigma$ with a measure $V^{-1}d\Sigma$ ($V\equiv \sqrt{-t^{\mu}t_{\mu}}$) is defined by
\begin{equation} \label{}
(\phi,\psi) \equiv  \int_{\Sigma}\phi^{*}\psi V^{-1}d\Sigma = \int^{\infty}_{-\infty} \phi^{*}\psi~ dx\,.
\end{equation}
Here, $\Sigma$ corresponds  to a Cauchy surface for a hyperbolic spacetime, and $V^{-1}d\Sigma$ is a certain measure which is realized in our case simply as $dx$. This inner product leads to a Hilbert space ${\cal H} = L^{2}((-\infty, \infty), dx )$.  

The Klein-Gordon equation in our background for the free theory in \eqref{Lfree} is given by
\begin{align}
        (\Box+m_0^2+\xi\mathcal{R})\phi=0\,,
\end{align}
which can be written as
\begin{align}
        \partial_t^2\phi=-A\phi\,,\qquad A=-\partial_x^2+e^{2\omega}(m_0^2+\xi\mathcal{R})\,.
\end{align}
As discussed in Section~\ref{SIFT}, SFTCS on  supersymmetric backgrounds can be interpreted as SIFT on flat background. Accordingly, the propagation of the scalar in the supersymmetric  curved background can be interpreted as that in flat background with position-dependent parameters. 
Concretely, the scalar wave equation in our background \eqref{susybackgd} is translated to the SIFT scalar wave equation with the position-dependent inhomogeneous mass-squared $m_{{\rm eff}}^2=e^{2\omega}(m_0^2+\xi\mathcal{R}) = m^2 + m'$ as
\begin{equation} \label{waveeq}
\partial^{2}_{t}\phi = -A \phi\,,  \qquad A = -\partial^{2}_{x} + m_{{\rm eff}}^{2}(x)\,,   
\end{equation}
where $A$ is a symmetric operator on the Hilbert space ${\cal H} = L^{2}((-\infty, \infty), dx )$.  

In~\cite{Ishibashi:2003jd}, the `energy' $E$ in curved spacetime is introduced by
\begin{equation} \label{}
E(\phi, \phi) = \frac{1}{2} \int^{\infty}_{-\infty} dx~  \dot{\phi}^{2}_{0} +  \frac{1}{2} \int^{\infty}_{-\infty} dx~  \phi_{0} A \phi_{0} \,,
\end{equation}
which is definitely  conserved. For the symmetric operator $A$ given in \eqref{waveeq}, the positivity of $E(\phi, \phi) $  is not guaranteed, which differs from \cite{Ishibashi:2003jd}. 
This difference is originated from the existence of the additional term $\xi \mathcal{R} \phi$ in \eqref{waveeq}. However, since our background metric is supersymmetric, the positivity of $E(\phi, \phi) $ holds.  This positivity of the  symmetric operator $ A$ can be shown by noting that  \begin{align}\label{A}
A =  D_{-}D_{+}, \qquad D_{\pm} \equiv \pm \frac{d}{dx} - m(x).
\end{align}
Then, for any $\phi \in \text{dom}(A)$, one obtains
\begin{equation} \label{}
(\phi, A\phi) =  \int^{\infty}_{-\infty} |D_{+}\phi|^{2}dx   \,,
\end{equation}
which  warrants the positivity of the operator $A$.   This is natural since the symmetric operator $A$ is identified with SQM Hamiltonian, and it is well-known that SQM Hamiltonian is non-negative~\cite{Cooper:1994eh}. Therefore,  this quantity $E(\phi, \phi) $ can be thought as an inner product of a `generalized' initial data space. It would be straightforward to check that all the requirements in \cite{Ishibashi:2003jd} are satisfied in our case. In fact, since our supersymmetric background in \eqref{susybackgd} is globally hyperbolic,  such a `generalization' is not necessary. 

Note that the above `energy' in the curved spacetime is related to the energy in IFT introduced through the canonical Hamiltonian $\sqrt{-g} \mathcal{H}_{\rm can}$ in \cite{Ho:2022ibc}. To compare the `energy', $E(\phi, \phi) $   in curved spacetime and the canonical energy in IFT, $E_{\rm IFT}$, we may note:
\begin{align}\label{Erel}
        E(\phi,\phi)=\int_{-\infty}^{+\infty} dx\,\bigg[ \sqrt{-g}\,\mathcal{H}_{\rm can}-\frac{1}{2}(\phi \phi')'\bigg]=E_{\rm IFT}-\frac{1}{2}\phi\phi'|_{-\infty}^{+\infty}\,.
\end{align}
Recall that the canonical Hamiltonian $\sqrt{-g} \mathcal{H}_{\rm can}$ was also used to define $E_{\rm BPS}$ in Sec.~\ref{25}. If we focus on the case that boundary term in \eqref{Erel} vanishes, $i.e.$ $\phi\phi'|_{-\infty}^{+\infty}=0$, we can see that $E(\phi,\phi)=E_{\rm IFT}$.

If the self-adjoint extension, $A_E$ of the symmetric operator $A$ in \eqref{waveeq} exists, a satisfactory dynamical evolution could be defined at least for initial data $(\phi_{0}, \dot{\phi}_{0})$ in $C^{\infty}_{0}(\Sigma)\times  C^{\infty}_{0}(\Sigma)$\footnote{This means that smooth initial data are compactly supported and the boundary condition, $\phi\phi'|_{-\infty}^{+\infty}=0$, is satisfied.} by
\begin{equation} \label{dynevol}
\phi_{t} = \cos(A^{1/2}_{E}t)\phi_{0} +   A^{-1/2}_{E} \sin(A^{1/2}_{E}t)\dot{\phi}_{0}\,,  \qquad (\phi_{0}, \dot{\phi}_{0}) \in {\cal H}\times {\cal H}\,.
\end{equation}
If more than one extensions of the symmetric operator $A$ exist,  these extensions lead to a non-unique dynamical evolution of the scalar field $\phi$. Essentially, different extensions would correspond to the different choice of boundary conditions at the singularity. However, as was shown   in~\cite{Ishibashi:2003jd}, the prescription for the dynamical evolution is determined uniquely  for the chosen extension $A_{E}$ under some specified requirements. 

On the other hand, we will show that $A_E$ is uniquely determined in our case,  so that the spacetime is globally hyperbolic. In other words,  the symmetric operator $A$ is essentially self-adjoint which means that the extension, $A_E$ is unique (See Ref.~\cite{Ishibashi:2004wx}.) and our naked null singularity is {\it mild}.  
 First, recall that $A = D_{-}D_{+}$ in \eqref{A}. In our supersymmetric background \eqref{susybackgd} $m(x)$ is written as \begin{equation} \label{G(x)}
m(x) = G^{-1}(x) \frac{dG(x)}{dx} \,, \qquad G(x) \equiv   (1+y)^{\beta}\,,~~{\rm or}~~ (1+y)^{1-\beta}\,,
\end{equation}
where we introduce a new variable  $y \equiv a e^{bx}$ and take $\beta\equiv2\xi=\frac{m_{0}}{ab}$.
 Using the time-like Killing symmetry of the system, one can take the separation of variable for the free scalar field in \eqref{waveeq}.
The relevant ordinary differential equation in our setup is 
\begin{equation} \label{ode}
A\phi_\omega(x) = \omega^{2} \phi_\omega(x) \,.
\end{equation}
To have the initial value problem well-posed, it is not enough that $A$ is a symmetric operator; it must be self-adjoint or allow self-adjoint extension. The method for finding the self-adjoint extension of a symmetric operator is well-known~\cite{Reed} and we will utilize it.
Briefly speaking, the essential self-adjointness of the operator $A$ can be tested by considering solutions of the equations
\begin{align}
	(A\pm i)\phi_{\omega}=0\,,
\end{align}
in the distributional sense. In our case, one can show that the operator $A$ is essentially self-adjoint since the square integrable solution to the equation~\eqref{ode} for $\omega^{2} =\pm i $ is trivial.
That is to say the {\it deficiency space} is empty~\cite{Ishibashi:2004wx}.
This will be explicitly shown in the next subsection.

\subsection{Analysis for scalar field dynamics}\label{42}

The equation \eqref{ode} can be regarded effectively as describing a quantum mechanical system where the operator $A$ plays the role of the Hamiltonian of the system:
%
\begin{equation} \label{waveeqation}
\bigg[ - \frac{d^{2} }{dx^{2}}  +  V_{\rm eff}(x) \bigg] \phi_{\omega}(x) =  \omega^{2}  \phi_{\omega}(x) \,,  \qquad   V_{\rm eff}(x) \equiv m_{{\rm eff}}^2(x)= \frac{(m_0^2e^{bx}+2 \xi ab^2)e^{bx}}{(ae^{bx}+1)^2 }\,.
\end{equation}
Here, without loss of generality, we can take $b>0$ as alluded in  Section \ref{SFTCS}. As discussed in Section \ref{SIFT}, the above quantum mechanical description for scalar field dynamics can be understood either in terms  of  SFTCS and SIFT. In the viewpoint of SFTCS,  the causal structure of the geometry is completely different according to the sign of $a$. This aspect is encoded in SIFT as the absence/presence of divergence of  $m^2(x)$.    Thus, we present separately effective potential forms for given parameters.      
     
     First of all, we define special points, $x=x_*, \bar x $, and $x_m$, by the conditions, 
\begin{align}
        V_{\rm eff}(x_*)=\pm\infty\,,\qquad V_{\rm eff}(\bar{x})=0\,,\qquad V_{\rm eff}'(x_{\rm m})=0,\,
\end{align}
which are given explicitly by 
\begin{align}\label{xsbm}
        x_{*}=\frac{1}{b}\log\Big(-\frac{1}{a}\Big)\,,\quad \bar{x}=\frac{1}{b}\log\Big(-\frac{2\xi}{a}\Big)\,,\quad x_{\rm m}=\frac{1}{b}\log\Big( \frac{1}{a(1-4\xi)}\Big)\,.
\end{align} 
We note that the value of $V_{{\rm eff}}$ at $x_m$ is given by
\begin{align} \label{maxminvalue}
        V_{\rm eff}(x_{\rm m})=\frac{b^2\xi}{2-4\xi}.
\end{align}

\begin{figure}[htbp]  
\begin{minipage}{0.48\textwidth}
\includegraphics[width=\linewidth]{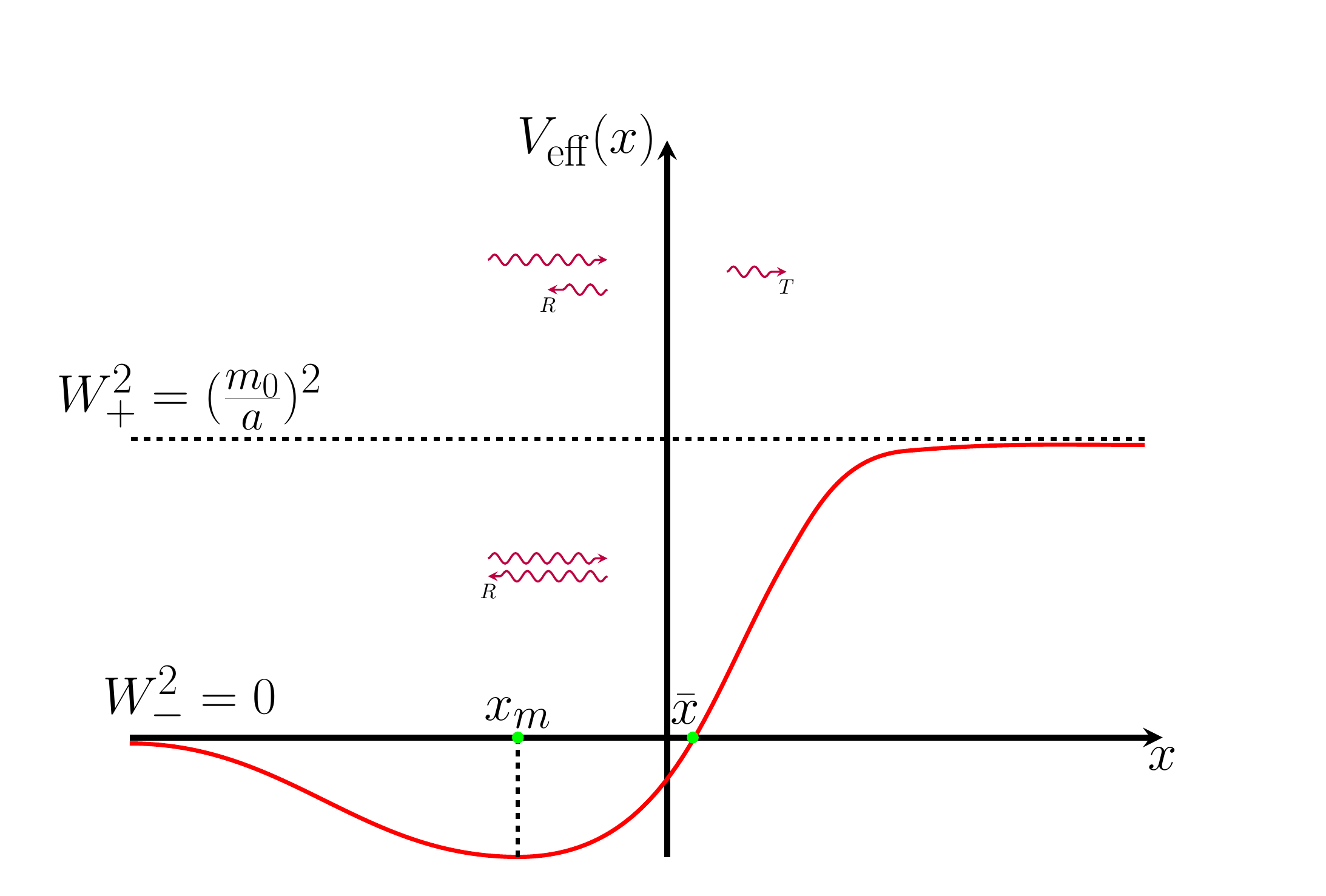}
\caption{$a>0$, $\xi <0 $ \\
$V_{\rm eff}$ has a minimum value given in \eqref{maxminvalue} at $x_{\rm m}$, and approaches to zero and $(m_0/a)^2$ as $x \rightarrow - \infty $ and  $x \rightarrow + \infty $, respectively.} 
\label{pa1}
\end{minipage} \hfill%
\begin{minipage}{0.48\textwidth}
\includegraphics[width=\linewidth]{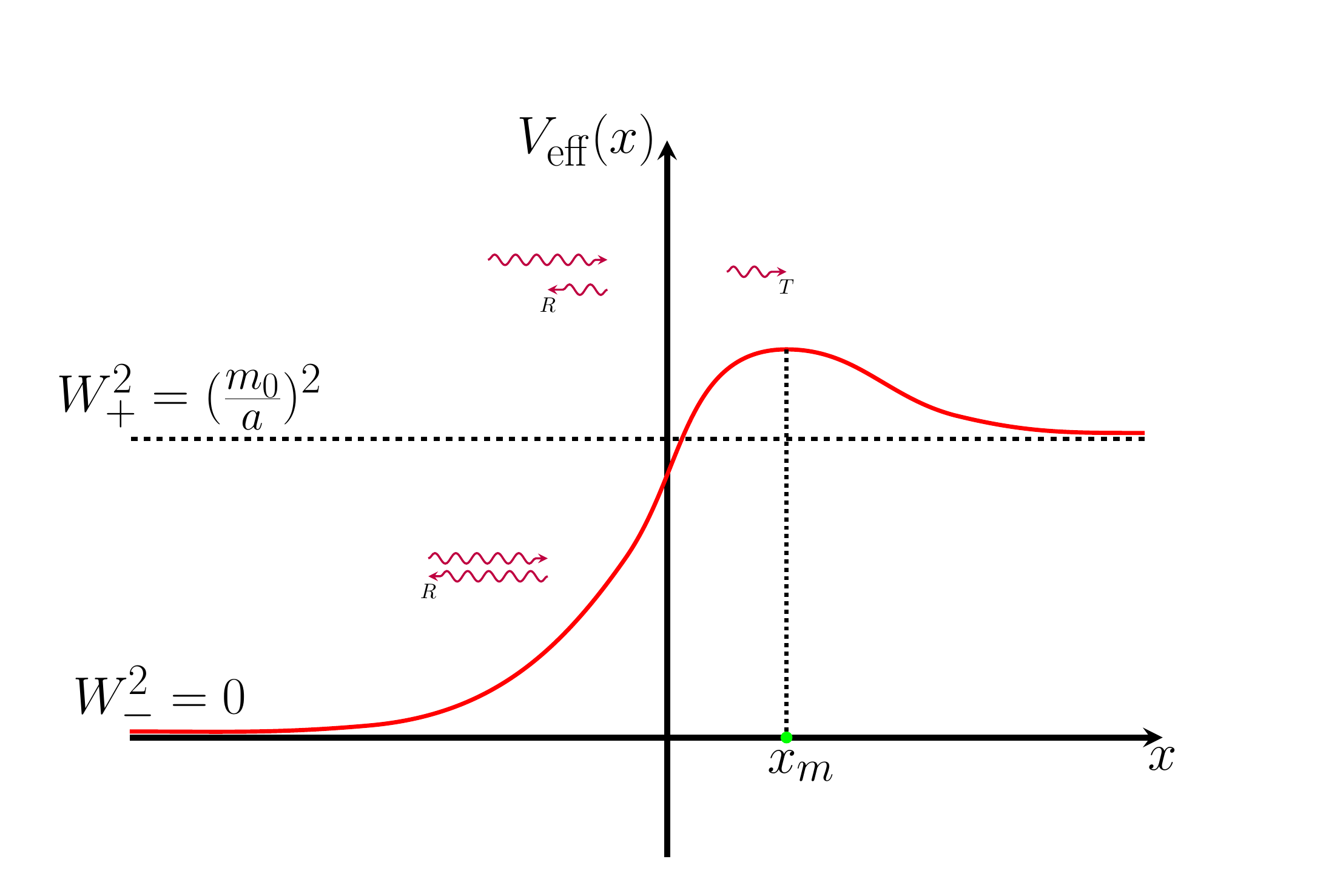}
\caption{$a>0$, $0<\xi < \frac{1}{4}$ \\
$V_{\rm eff}$ has a maximum value given in \eqref{maxminvalue} at $x_{\rm m}$, and approaches to zero and $(m_0/a)^2$ as $x \rightarrow - \infty $ and  $x \rightarrow + \infty $, respectively.}
\label{pa2}
 \end{minipage}
\label{PAA}
 \end{figure}

%
%

\begin{figure}[htbp]  
\begin{minipage}{0.48\textwidth}
\includegraphics[width=\linewidth]{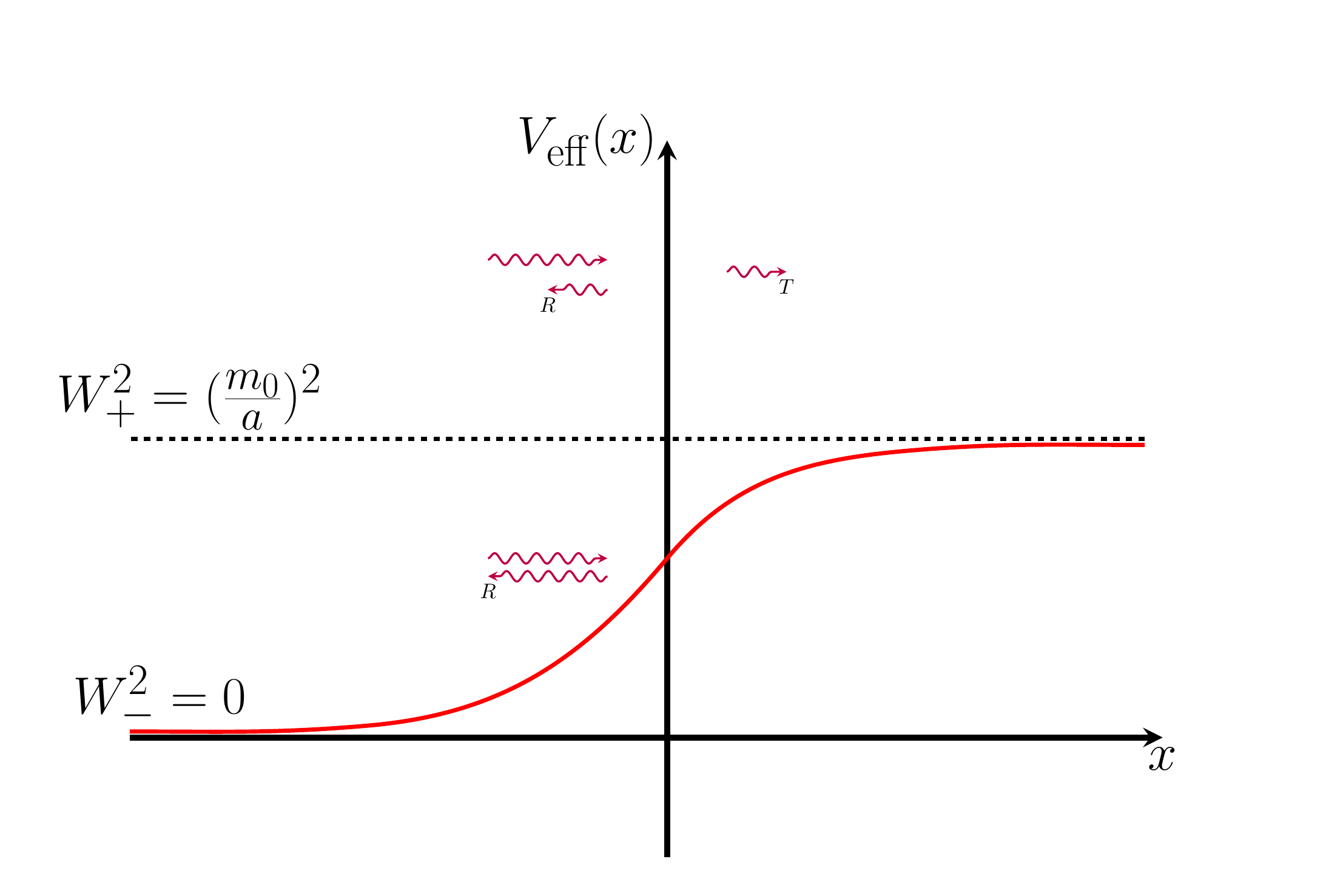}
\caption{$a>0$, $\xi \ge \frac{1}{4}$ \\
$V_{\rm eff}$ has a $S$-shaped graph without any minimum/maximum values, and approaches to zero and $(m_0/a)^2$ as $x \rightarrow - \infty $ and  $x \rightarrow + \infty $, respectively.}
\label{pa3}
\end{minipage} \hfill%
\begin{minipage}{0.48\textwidth}
\includegraphics[width=\linewidth]{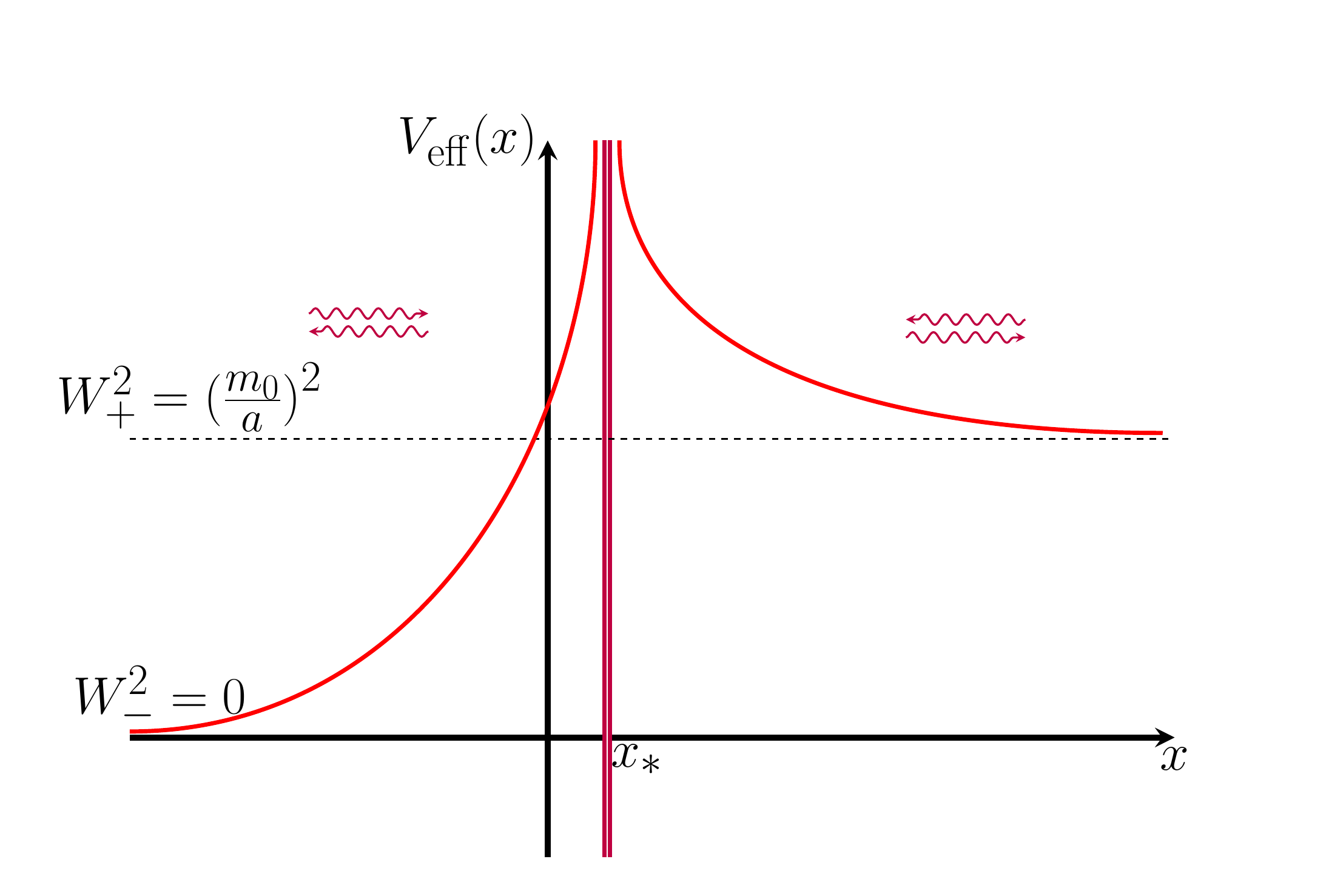}
\caption{$a<0$, $\xi <0 $ \\
$V_{\rm eff}$ diverges to $+ \infty $ at $x_{*}$, and approaches to zero and $(m_0/a)^2$ as $x \rightarrow - \infty $ and  $x \rightarrow + \infty $, respectively.}
\label{na1}
 \end{minipage}
\label{PBB}
 \end{figure}

%


\begin{figure}[htbp]  

\begin{minipage}{0.48\textwidth}
\includegraphics[width=\linewidth]{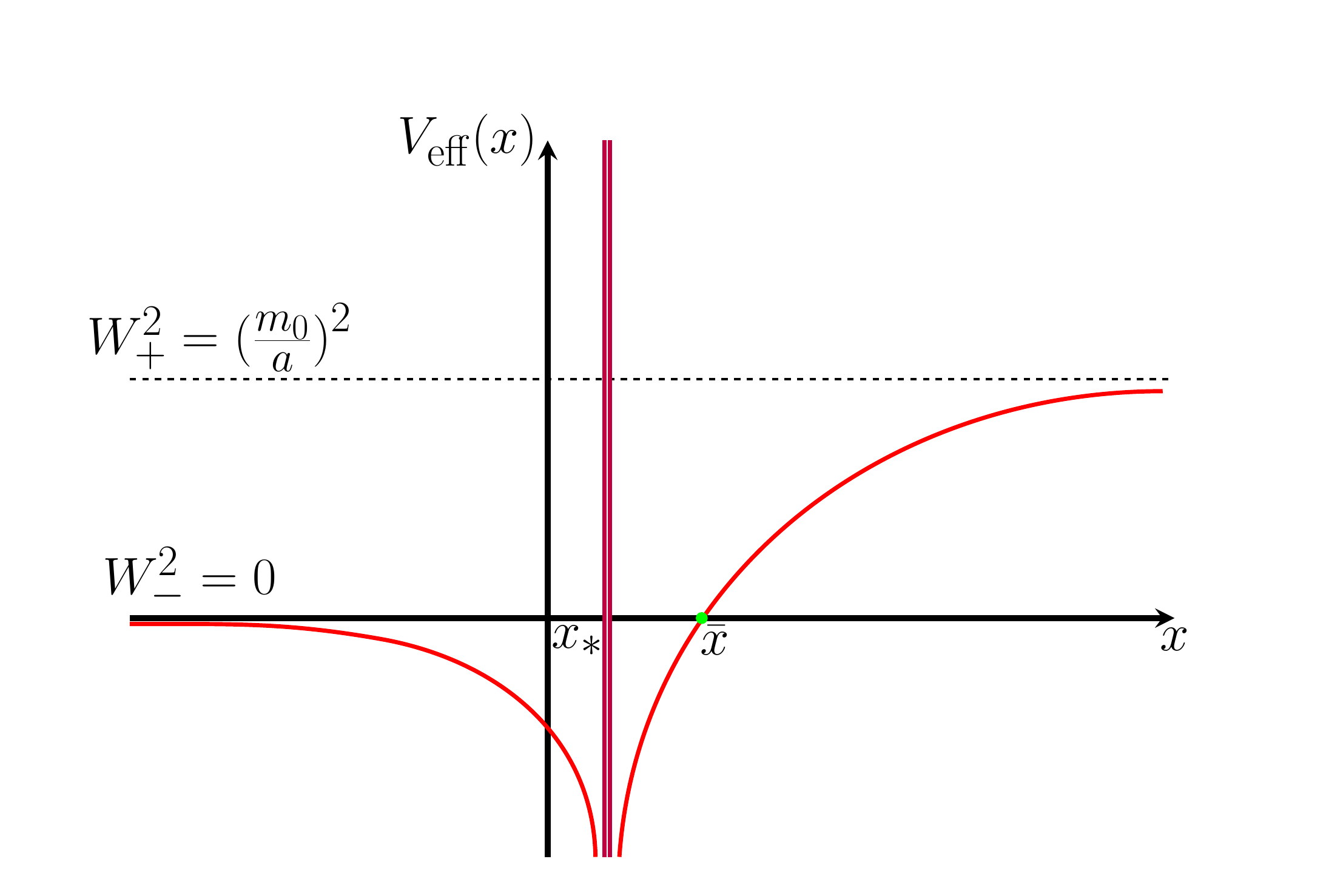}
\caption{$a<0$, $0<\xi \leq \frac{1}{4} $ \\
$V_{\rm eff}$ diverges to $- \infty $ at $x_{*}$, and approaches to zero and $(m_0/a)^2$ as $x \rightarrow - \infty $ and  $x \rightarrow + \infty $, respectively.}
\label{na3}
 \end{minipage}\hfill%
 \begin{minipage}{0.48\textwidth}
\includegraphics[width=\linewidth]{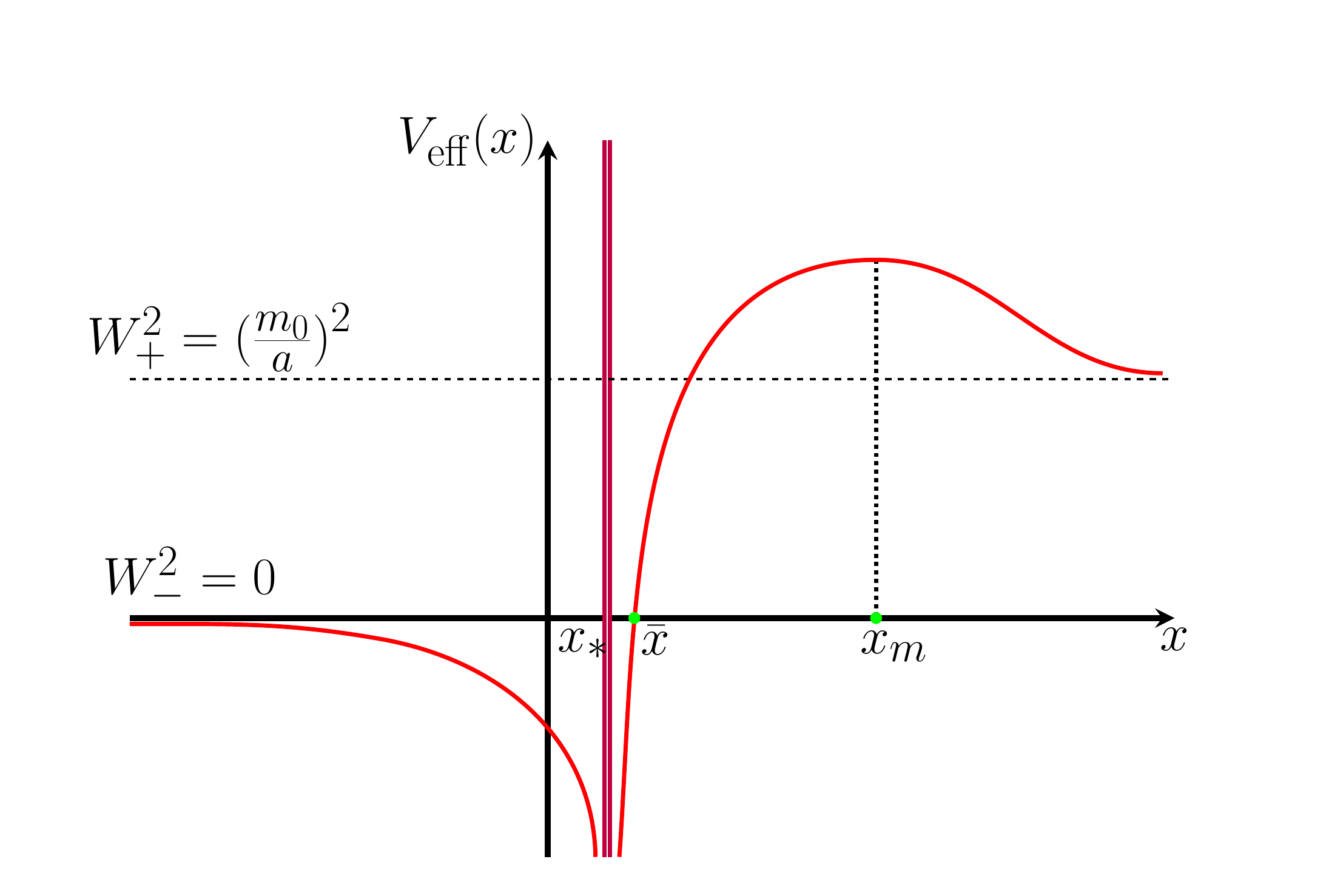}
\caption{$a<0$, $\frac{1}{4} < \xi < \frac{1}{2} $ \\
$V_{\rm eff}$ has a maximum value given in \eqref{maxminvalue} at $x_{\rm m}$, and approaches to zero and $(m_0/a)^2$ as $x \rightarrow - \infty $ and  $x \rightarrow + \infty $, respectively. It diverges to $ - \infty$ at $x_*$.}
\label{na4}
\end{minipage} 
\label{PCC}
 \end{figure}

%
%

\begin{figure}[htbp]  
\begin{minipage}{0.48\textwidth}
\includegraphics[width=\linewidth]{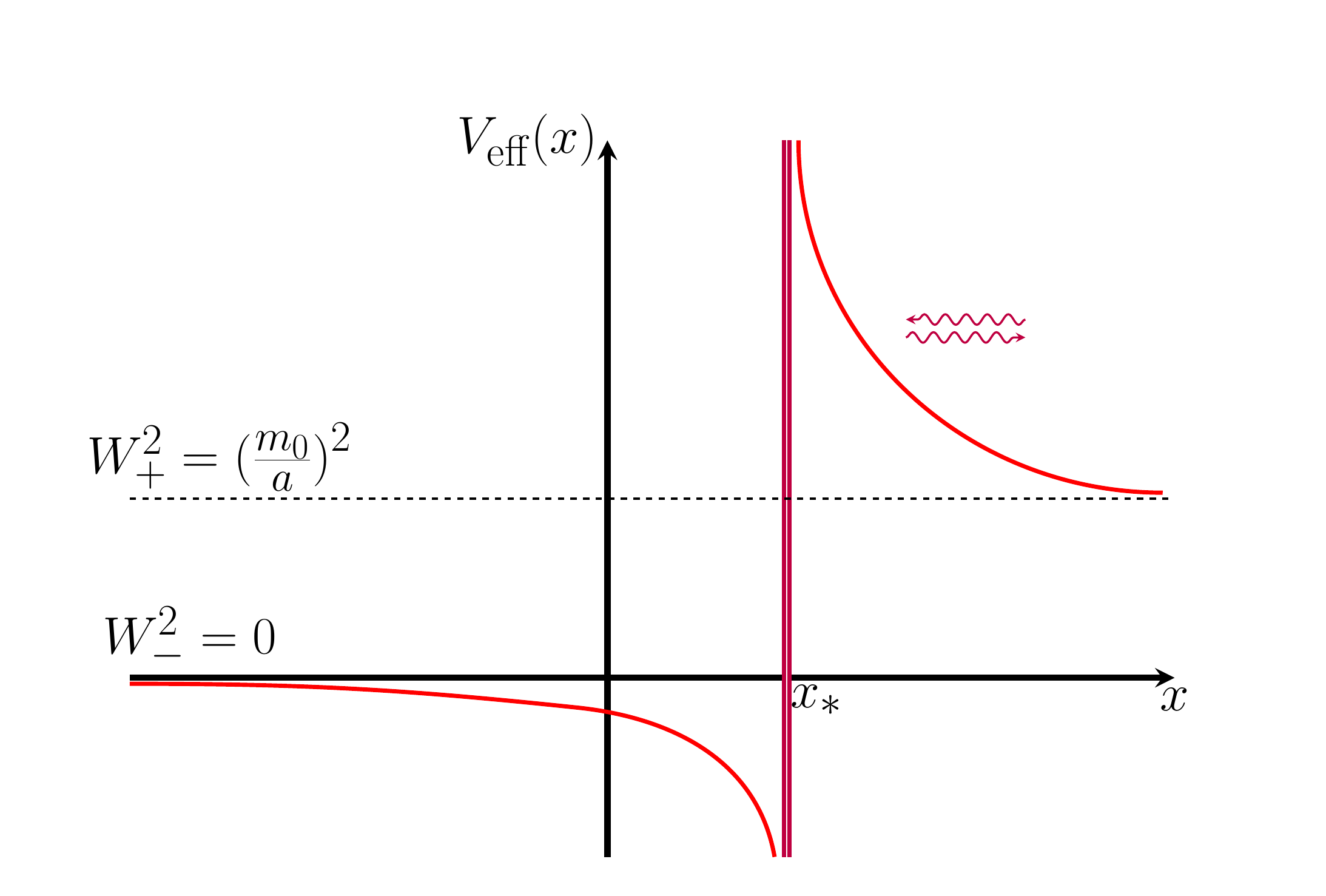}
\caption{$a<0$, $\xi =\frac{1}{2} $ \\
$\lim_{x \to x_* \pm} V_{\rm eff} = \pm \infty $. $V_{\rm eff}$ approaches to zero and $(m_0/a)^2$ as $x \rightarrow - \infty $ and  $x \rightarrow + \infty $, respectively.}
\label{na2}
\end{minipage} \hfill%
\begin{minipage}{0.48\textwidth}
\includegraphics[width=\linewidth]{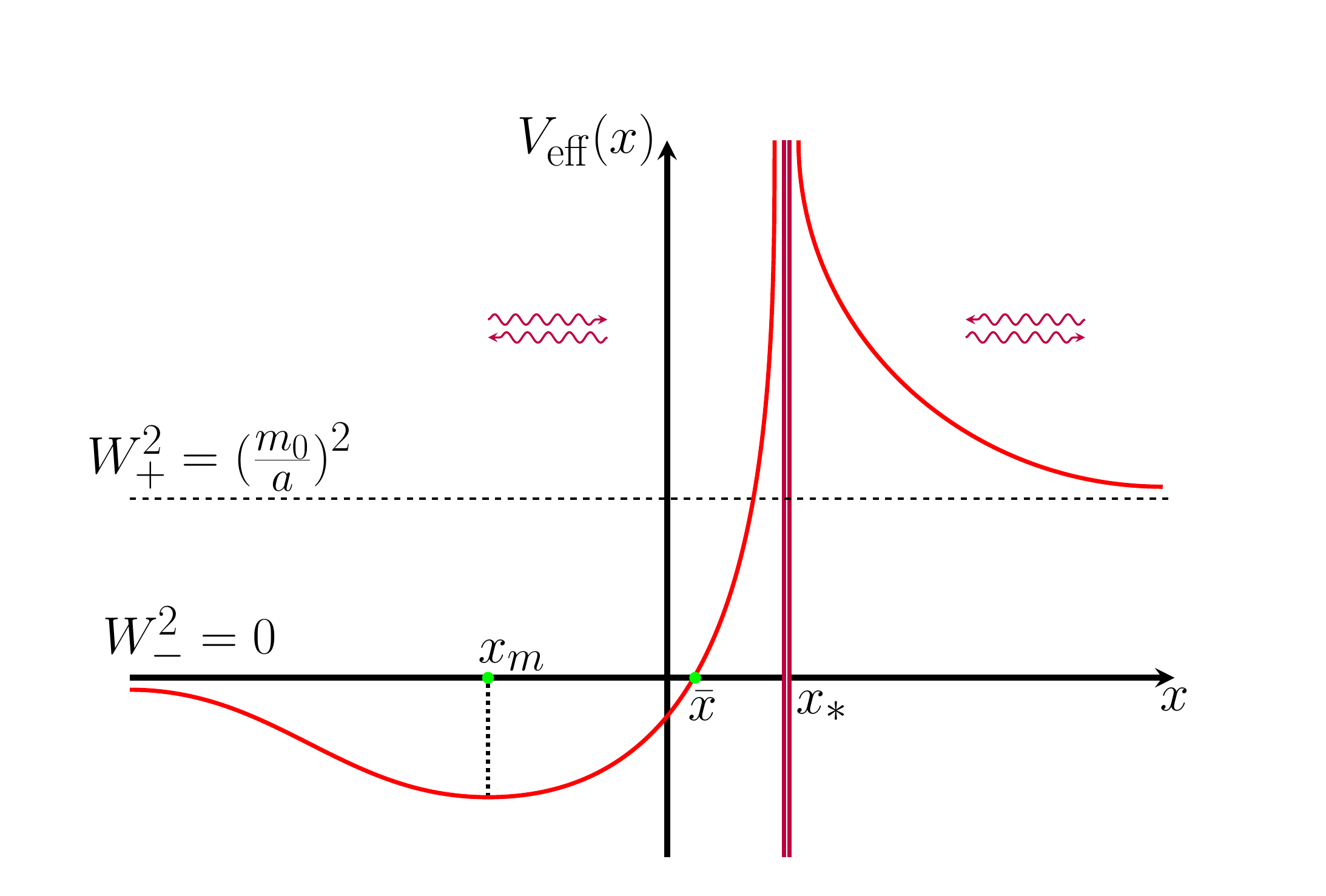}
\caption{$a<0$, $\xi > \frac{1}{2} $ \\
$\lim_{x \to x_* } V_{\rm eff} =  \infty $. $V_{\rm eff}$ approaches to zero and $(m_0/a)^2$ as $x \rightarrow - \infty $ and  $x \rightarrow + \infty $, respectively. It has a minimum value given in \eqref{maxminvalue} at $x_{\rm m}$.}
\label{na5}
 \end{minipage}
\label{PDD}
 \end{figure}

%
%
%
%

According to the sign of $a$, the profiles of $V_{{\rm eff}}$ are quite different. We sketch the characteristics briefly in Fig.~\ref{pa1} - Fig.~\ref{na5} :

\noindent 
$(i)$ $a>0$ : Depending on the range of $\xi$, $V_{{\rm eff}}$ has a minimum and a maximum values at $x_m$ for $\xi <0$ and  $0< \xi < \frac{1}{4}$, respectively. Rather, it takes a monotonic behavior for  $\xi \ge \frac{1}{4}$. See Fig.~\ref{pa1} - Fig.~\ref{pa3}.

\noindent 
$(ii)$ $a<0$ : The spacetime consists of causally disconnected L and R parts. This fact is reflected in $V_{\rm eff}(x\rightarrow x_{*})\rightarrow \pm\infty$.  We have plotted two different systems, L and R parts, in a single diagram, just like the corresponding Penrose diagram in Fig.~\ref{PenroseD}($b$). The profiles of $V_{\rm eff}$ are given in  Fig.~\ref{na1} - Fig.~\ref{na5}. 
For $\frac{1}{4}<\xi<\frac{1}{2}$, there exists a local maximum value of $V_{\rm eff}$ at $x_{\rm m}$ in R part 
(Fig.~\ref{na3}).  On the contrary, for $\xi>\frac{1}{2}$, $V_{\rm eff}$ has a negative minimum value at $x_{\rm m}$ in L part (Fig.~\ref{na5}).

\noindent 
$(iii)$ $a=0$ : This case corresponds to Rindler spacetime and, as is well-known,  the profile of $V_{\rm eff}$ is  exponentially increasing, $V_{\rm eff}^{a=0}(x) = m_0^2\, e^{2bx}$.  We do not discuss this case further and relegate detalils to \cite{Takagi:1986kn, Crispino:2007eb, Ho:2022ibc}.

Recall that the metric parameters $a$ and $b$ are not independent since  those are related to the given Lagrangian parameters $m_{0}$ and  $\xi$ as  $2ab \xi =m_{0}$. Using this relation, we rewrite the effective potential $V_{\rm eff}$ as  
\begin{align}
        V_{\rm eff}^{a>0}(x) &=2b^{2}\xi^{2}\Big[1  +  \tanh\frac{b}{2}(x-x_{0}) \Big] - \frac{b^{2}\xi}{2}\Big(2\xi-1\Big)\frac{1}{\cosh^{2}\frac{b}{2}(x-x_{0})}\,, \label{Veff1}\\
        V_{\rm eff}^{a<0}(x) &= 2b^{2}\xi^{2}\Big[1  +  \coth\frac{b}{2}(x-x_{*}) \Big] + \frac{b^{2}\xi}{2}\Big(2\xi-1\Big)\frac{1}{\sinh^{2}\frac{b}{2}(x-x_{*})}\,, \label{Veff3}
\end{align}
where $x_0$ is defined by $e^{-bx_{0}} \equiv a$ and $x_*$ was introduced in \eqref{xsbm}.

Note that the potential \eqref{Veff1} belongs to the category of hyperbolic Rosen-Morse potential, and the potential \eqref{Veff3} does to that of Eckart potential.\!\footnote{Our potentials correspond to taking $\alpha=\pm\frac{b}{2},\, A=\pm b\xi$, and $B=\mp b^2\xi^2$ for the parameters  $\alpha$, $A$, and $B$ in Table 1,2 of \cite{Khare:1988tg}.} In the case of $a<0$, there are two parts alluded before, L and R parts. To confirm that our potential \eqref{Veff3} belongs to the standard Eckart potential form, one can set $r\equiv x-x_*\geq0$ in the R part, and $r\equiv-(x-x_*)\geq0$ in the L part.

Now, let us remind that the symmetric operator $A$ is factorized as in  \eqref{A} and corresponds to the Hamiltonian of SQM. In our case, the superpotential is given by 
\begin{equation} \label{QMSP1}
W_{\rm QM}(x) = -G^{-1}(x) \frac{d}{dx}G(x) = - m(x) \,,
\end{equation}
which satisfies $V_1 \equiv V_{\rm eff} =W_{\rm QM}^{2} - \frac{dW_{\rm QM}}{dx}$.
Then, the super partner potential of $V_{1}(x)$ is given by
\begin{equation} \label{}
V_{2}(x) = W_{\rm QM}^{2} + \frac{dW_{\rm QM}}{dx}\,.
\end{equation}
Note that $G(x)$ corresponds to 
the ground state wave function in standard SQM~\cite{Cooper:1994eh} when the bound states exist. However, when bound states do not exist, we cannot identify $G(x)$ with the ground state wave function, since $G(x)$ is not square integrable. In the following, our main concern is a non-square integrable function $G(x)$ in \eqref{G(x)}. As was shown before, bound states may exist for some ranges of $\xi$ and $a$. In this case, $G(x)$ may be interpreted as a ground state wave function.

For our background,  by using $G(x)= (1+ a e^{bx})^\beta$  the superpotential $W_{\rm QM}(x)$ becomes
\begin{align}\label{WQM}
        W_{\rm QM}(x) &= - 2b\xi \frac{a e^{bx}}{1+ae^{bx}}=
        \begin{cases}
                & -b\xi\Big[1+\tanh\frac{b}{2}(x-x_{0})\Big]\,,\qquad a>0\,, \\
                & -b\xi\Big[1+\coth\frac{b}{2}(x-x_{*})\Big]\,,\qquad a<0\,.
        \end{cases}
\end{align}
 Furthermore, we can check that the potential $V_{\rm eff}(x) = V_1(x)$  has the so-called shape-invariance. To see this, one may compute $V_2$ as
\begin{align}\label{V2}
        V_{2}(x) &=
        \begin{cases}
                & 2b^{2}\xi^{2}\Big[1  +  \tanh\frac{b}{2}(x-x_{0}) \Big] - \frac{b^{2}\xi}{2}\Big(2\xi+1\Big)\frac{1}{\cosh^{2}\frac{b}{2}(x-x_{0})}\,,\qquad a>0\,, \\
                & 2b^{2}\xi^{2}\Big[1  +  \coth\frac{b}{2}(x-x_{*}) \Big] + \frac{b^{2}\xi}{2}\Big(2\xi+1\Big)\frac{1}{\sinh^{2}\frac{b}{2}(x-x_{*})}\,,\qquad a<0\,.
        \end{cases}
\end{align}
One can easily see  the shape invariance of the above the Rosen-Morse/Eckart potentials by noting that $V_{2}(x\,;\,-\xi) = V_{\rm eff}(x\,;\, \xi)$.
This corresponds to a special case of the shape invariance of the Rosen-Morse/Eckart potentials in \cite{Khare:1988tg}.

For clear presentation, we will focus on the case of $a>0$ and $\xi\geq\frac{1}{4}$ in Fig.~\ref{pa3}.
Though this potential is known to be exactly solvable and studied in some contexts, the concrete profile of the potential in our case is slightly different in the sense that there is no bound state in the view point of quantum mechanics. That is to say, the potential is monotone   and the minimum of the potential is located at $x\rightarrow -\infty$.

As a next step, we present some details on the solution to the scalar wave equation for completeness of our discussion. 
By taking the change of variable $y = a e^{bx} = e^{b(x-x_{0})}$ in \eqref{waveeqation} and setting 
\begin{equation} \label{alphabeta}
\phi_{\omega}(y) \equiv y^{\alpha}(1+y)^{\gamma}f_{\omega}(y)
\end{equation}
with $\alpha=\pm i\frac{\omega}{b}$ and $\gamma=\beta$ or $\gamma=1-\beta$, one can see that the differential equation satisfied by $f_{\omega}(y)$ becomes a hypergeometric differential equation in the form of 
\begin{equation} \label{}
\Big[y(1+y) \frac{d^{2}}{dy^{2}}   + \Big( 2\alpha +1 + (2\gamma + 2\alpha +1) y\Big) \frac{d}{dy} + \gamma(2\alpha+1)-2\xi \Big]  f_{\omega} (y)  =0\,.
\end{equation}
Then, we can immediately write  down solutions to our  differential equations in terms of hypergeometric functions as 
\begin{equation} \label{phisol}
\phi_{\omega}(y)  =  (1+y)^{\gamma} \Big[ a_{1} y^{\alpha}F(A,B\,;\, C\,|\, -y) + a_{2}~ y^{\alpha+1-C}F(A-C+1,B-C+1\,;\, 2-C\,|\, -y) \Big]\,,    
\end{equation}
where  $A,B,C$ are given by $A+B = 2 \alpha + 2\gamma,\, AB = 2\alpha\gamma,\,  C= 1 + 2\alpha$.
Now,  we can take  the parameters $\alpha=\frac{i}{b} \omega$ and $\gamma=\beta$.
To see the generality in the choice of $\alpha$, note that two hypergeometric functions in~\eqref{phisol} with the front $y$ factors interchange by $\alpha\rightarrow -\alpha$. 
For the generality in the choice of $\gamma$, apply the hypergeometric function transformation property to each hypergeometric function respectively in~\eqref{phisol}
\begin{equation} \label{}
F(A,B\,;\, C\,|\, z) = (1-z)^{C-A-B} F(C-A,C-B\,;\, C\,|\, z)\,.
\end{equation}
Hence, without loss of generality,  we can take  
\begin{equation} \label{ABC}
 A = \frac{i}{b} (\omega - k)  + \beta \,, \qquad  B = \frac{i}{b} (\omega  + k) + \beta\,, \qquad C = 1 + 2\frac{i}{b}\omega\,,
\end{equation}
where  $k$ is defined by 
\begin{equation} \label{}
k^{2} \equiv  \omega^{2} - (2b\xi)^{2}\,.
\end{equation}
Note that this choice implies  $A-B = -2i\frac{k}{b}$.

There are three cases for the allowed values of $k^{2}$ in the above: {\bf Case $(i)$} $k^{2} >  0$, {\bf Case $(ii)$}  $k^{2} < 0$, and {\bf Case $(iii)$}  $k^{2} =0$. These cases need to be treated separately. 

In {\bf Case $(i)$}, by using the linear transformation of hypergeometric function given  in~\eqref{HyperP1}, one can rewrite  the solution in the form of 
\begin{align}    \label{phiw}
\phi_{\omega}(y) &=     (1+y)^{\beta} \Big[ b_{1}~y^{\alpha-A} F\Big(A,A-C+1\,;\, A-B+1\,\Big|\, -\frac{1}{y} \Big)  \nonumber \\ 
& \qquad \qquad \qquad  \qquad + b_{2}~ y^{\alpha-B}F\Big(B,B-C+1\,;\, B-A+1\,\Big|\, - \frac{1}{y} \Big) \Big]\,, 
\end{align}
where the constants $b_{1,2}$ (with  $\beta \in {\bf R}$) are related to the previous constants $a_{1,2}$ as follows
\begin{align}    \label{}
b_{1} &= \frac{\Gamma(2i\frac{k}{b})\Gamma(1+2i\frac{\omega}{b}) [a_{1} - R^{*}_{\omega}\, a_{2}]  }{\tiny \Gamma (\beta + \frac{i}{b}(\omega+k) )\Gamma (1-\beta + \frac{i}{b}(\omega+k) )}\,, \nonumber \\
b_{2} &=  \frac{ \Gamma(-2i\frac{k}{b})\Gamma(1-2i\frac{\omega}{b}) [-R_{\omega} a_{1} + a_{2} ]}{\tiny \Gamma(\beta - \frac{i}{b}(\omega+k))\Gamma (1-\beta - \frac{i}{b}(\omega+k))} \,. \label{connection}
\end{align}
Here,  $R_{\omega} \in {\bf C}$ is  defined by
\begin{equation} \label{RefC0}
R_{\omega} \equiv  -{\textstyle \frac{\Gamma(C)\Gamma(A-C+1)\Gamma(1-B)}{\Gamma(2-C)\Gamma(A)\Gamma(C-B)} }=  -{ \textstyle  \frac{\Gamma\big(1+2i\frac{\omega}{b}\big)\Gamma\big(\beta - \frac{i}{b}(\omega+k)\big)\Gamma \big(1-\beta - \frac{i}{b}(\omega+k)\big)}{\Gamma\big(1-2i\frac{\omega}{b}\big)\Gamma\big(\beta + \frac{i}{b}(\omega-k)\big)\Gamma \big(1-\beta  + \frac{i}{b}(\omega-k)\big)}  }\,.
\end{equation}
Introducing $\theta_{k}, \bar{\theta} \in {\bf R} $ as 
\begin{equation} \label{}
e^{i\theta_{k}} \equiv \frac{\Gamma(\beta - \frac{i}{b}(\omega + k))}{ \Gamma(\beta + \frac{i}{b}(\omega + k))} \frac{\Gamma(1-\beta - \frac{i}{b}(\omega + k))}{ \Gamma(1-\beta + \frac{i}{b}(\omega + k))}\,, \qquad  e^{i\bar{\theta}} \equiv \frac{\Gamma\big(1+2i\frac{\omega}{b}\big)}{\Gamma\big(1-2i\frac{\omega}{b}\big)}\,,
\end{equation}
one can rewrite $R_{\omega}$ as
\begin{equation} \label{RefC} 
 R_{\omega}  
 = - e^{\frac{i}{2} (\theta_{k}+\theta_{-k} +2\bar{\theta})} {\textstyle \sqrt{ \frac{\sin^{2}\pi\beta + \sinh^{2}\frac{\pi}{b}(\omega -k)}{\sin^{2}\pi\beta + \sinh^{2}\frac{\pi}{b}(\omega + k)}} } \,, \qquad
R^{*}_{\omega} = \frac{1}{R_{-\omega}}\,.
\end{equation}
The solution form of \eqref{phiw} is useful  to see the behavior at the asymptotic region $x\rightarrow \infty$ ($y\rightarrow\infty$) and its leading asymptotic behavior is given by  
\begin{equation} \label{phiw1}
\phi_{\omega}(x) \underset{x\rightarrow\infty}{\longrightarrow} b_{1}e^{ik(x-x_{0})} + b_{2} e^{-ik(x-x_{0})}\,,
\end{equation}
 where we have used~\eqref{ABC} to obtain $\alpha+\beta -A = \frac{i}{b}k$ and $\alpha+\beta -B = -\frac{i}{b}k$. On the contrary, in the asymptotic region of $x\rightarrow -\infty$ ($y\rightarrow 0$), one can read from \eqref{phisol} that
\begin{equation} \label{phiw2}
\phi_{\omega}(x)  \underset{x\rightarrow -\infty} \longrightarrow a_{1} \, e^{i\omega (x-x_{0})} + a_{2} \, e^{-i\omega (x-x_{0})}\,.
\end{equation}
As explained after \eqref{ode}, it is necessary to verify the essential self-adjointness of the symmetric operator $A$ for sensible dynamics. Therefore, we should extend the domain of $\omega^2$ from real to complex numbers and consider the case of $\omega^2=\pm i$.
Now, we show that a square-integrable solution to the equation \eqref{ode} for $\omega^2=\pm i$ does not exist. The square-integrability of $\phi_\omega(x)$ implies the condition $a_1 a_2=0$. From the relation $k^2=\omega^2-b^2\beta^2$ and square-integrability, the condition $b_1 b_2=0$ should be satisfied. Then the connection formulae \eqref{connection} with these conditions tell us that square integrable solution for $\omega^2=\pm i$ does not exist, indeed.
So, the operator $A$ is essentially self-adjoint and the initial value problem of wave dynamics is well-posed.

In {\bf Case $(ii)$},  let us set $k=-i\kappa$ with  $\kappa >0$ without loss of generality. Note that $\tilde{R}_{\omega} \equiv  R_{\omega}(k=-i\kappa)$ is a pure phase and corresponds to a reflection coefficient in this case.
Let us consider $\frac{2\kappa}{b} \notin{\bf N}$, firstly. Then, we should take $b_{1} =0$ for a square integrable solution, which implies $a_{1} = R^{*}_{\omega}a_{2}$. Secondly, in the case of $\frac{2\kappa}{b} \in {\bf N}$, we need to use a more complicated formula given in~\eqref{HyperP2}, instead of the formula in~\eqref{HyperP1}, and obtain
\begin{equation} \label{phiwc}
\phi_{\omega}(x)  \underset{x\rightarrow\infty}{\longrightarrow} \sum_{n=0}^{\frac{\kappa}{b}-1}  c_{1, n}\,  e^{b(\frac{\kappa}{b} -n) x}       + c_{2}\, x\,  e^{-\kappa x} + \cdots\,,
\end{equation}
where $\cdots$ denotes lower orders in $e^{-\kappa x}$. Now,  for the square integrable solution, it is sufficient to set $a_{1} = \tilde{R}^{*}_{\omega} a_{2}$. 
  The square-integrability of $\phi_\omega(x)$, as can read from \eqref{phiw2}, implies the condition $a_1 a_2=0$. Since $a_{1} \propto a_2$ in both cases, we  conclude that a square-integrable solution to the equation \eqref{ode} for $\omega^2=\pm i$ does not exist.

In {\bf Case $(iii)$},  we can still use the formula \eqref{HyperP2}. By using $k=\kappa=0$, one obtains 
\begin{equation} \label{}
\phi_{\omega}(x)  \underset{x\rightarrow\infty}{\longrightarrow}   c_{3}  + c_{4}x\,.
\end{equation}
To obtain a  meaningful solution, we should choose $c_3=c_{4} =0$ for square integrability.

In all three cases, a square-integrable solution to the equation \eqref{ode} with $\omega^2=\pm i$ does not exist, which means 
 the emptiness of the {\it deficiency space}. Consequently, the positive symmetric operator $A$ is essentially self-adjoint, 
and so the wave propagation in this spacetime with naked null singularity is well-defined and any boundary condition is  not required at singularity, contrary to the timelike singularity in \cite{Wald:1980jn}. In other words, the singularity is a {\it mild} one. Though such a description has been given from SFTCS viewpoint,  
the absence of boundary condition at singularity is natural from SIFT viewpoint, since $m^2(x)$ in SIFT has no unusual profile.

\subsection{Scalar field quantization}\label{AppD}

In this subsection, we explore the canonical quantization of a scalar field with a potential, as depicted in Fig.~\ref{pa3}.  We discuss the limitations of this approach, placing special emphasis on the algebraic method.

To address canonical quantization, we note that our mode solution can be expressed as two separate plane waves in the right and left asymptotic regions, as demonstrated by equations \eqref{phiw1} and \eqref{phiw2}, respectively. Based on this observation, it is natural to consider two  canonical quantization schemes that turn out to be inequivalent. In the following, we refer to the quantization scheme that uses the mode solution presented in equation \eqref{phiw1} as R-quantization, while the scheme that employs the mode solution given in equation \eqref{phiw2} is referred to as L-quantization.  We would like to emphasize that the FTCS and IFT viewpoints can be used in a complementary manner to interpret the results of quantization. It  is evident that both FTCS and IFT descriptions are entirely interchangeable at the classical level from the equivalence of the classical scalar field equation. At the quantum level, we adopt our proposal in~\cite{Ho:2022ibc}, and then the interpretation becomes complementary.

  First, in L-quantization, from the classical  solution in~\eqref{phiw2}, we canonically quantize the scalar field as 
\begin{align}\label{phiL}
	\phi_{\rm L}(\pmb{x}) &= \int_{0}^{\infty}\frac{d\omega}{\sqrt{2\pi}} \frac{1}{\sqrt{2\omega}}\sum_{i=\pm}
\bigg[ a_{\omega}^{(i)}u^{(i)}_{\omega}(\pmb{x}) +\big(a_{\omega}^{(i)}\big)^{\dagger}\big(u^{(i)}_{\omega}(\pmb{x})\big)^{*}  \bigg]\,, 
\end{align}
where $u^{(\mp)}_{\omega}(\pmb{x})$'s are given by\footnote{Here, \eqref{up} and \eqref{um} have been read from \eqref{phisol} and $a=e^{-bx_0}$ is used. Then, we denote $x-x_{0}$ simply as $x$ to simplify the presentation. The normalization is determined through the Klein-Gordon inner product. }
\begin{align}
u^{(-)}_{\omega}(\pmb{x}) &= (1+ e^{bx})^{\beta}   F(A,B\,;\, C\,|\, - e^{bx}) e^{-i\omega (t-x)} \,,    \label{um} \\
	u^{(+)}_{\omega}(\pmb{x}) &=   (1+e^{bx})^{\beta} F(A-C+1,B-C+1\,;\, 2-C\,|\, - e^{bx}) e^{-i\omega (t+x)}\,,\label{up} 
\end{align}
and complex conjugates of mode functions are given by
\begin{align}
 \big(u^{(-)}_{\omega}(\pmb{x})\big)^{*} &=  (1+ e^{bx})^{2\xi} F(A-C+1,B-C+1\,;\, 2-C\,|\, - e^{bx}) e^{i\omega (t-x)} \,,   \label{umcc} \\
	\big(u^{(+)}_{\omega}(\pmb{x})\big)^{*} &=  (1+e^{bx})^{2\xi}   F(A,B\,;\, C\,|\, - e^{bx}) e^{i\omega (t+x)}\,. \label{upcc}
\end{align}
Note that
$\big[F(A,B\,;\, C\,|\, - e^{bx})\big]^{*}=F(A-C+1,B-C+1\,;\, 2-C\,|\, - e^{bx})$
as seen from \eqref{ABC}.
Our normalization convention for $u^{(\mp)}_{\omega}(\pmb{x})$ is in accordance with the non-vanishing commutator of creation and annihilation operators, which can be expressed as follows
\begin{equation} \label{aCom}
[a_{\omega}^{(i)}, (a_{\omega'}^{(j)}\big)^{\dagger}]  = \delta^{ij} \delta(\omega -\omega')\,,
\end{equation}
where $i$ and $j$ denote $-$ or $+$. In fact, the operators $a^{(-)}_{\omega}$ and $a^{(+)}_{\omega}$ are not independent for the range of $ 0  \le \omega \le b\beta$, as we will discuss shortly. We define the vacuum in L-quantization, $|0\rangle_{\rm L}$, as the state annihilated by $a_{\omega}^{(\mp)}$ as
\begin{equation} \label{}
a_{\omega}^{(\mp)}|0\rangle_{\rm L}=0\,.
\end{equation}
The Fock space constructed by $a^{(\mp)}_{\omega}$ and $\big(a^{(\mp)}_{\omega}\big)^{\dagger}$ is denoted by $\mathscr{F}_{\rm L}$.
At the left asymptotic region, $u^{(\mp)}_{\omega}(\pmb{x})$'s reduce to
\begin{align}
	u^{(\mp)}_{\omega}(\pmb{x}) \underset{x\rightarrow -\infty}\longrightarrow e^{-i\omega(t\mp x)}\,,
\end{align}
and in L-quantization scheme, the form of scalar field at the left asymptotic region $(x\rightarrow-\infty)$ approaches to
\begin{equation} \label{}
\phi_{\rm L} (\pmb{x}) \underset{x\rightarrow-\infty}{\simeq} \int^{\infty}_{0} \frac{d\omega}{\sqrt{2\pi}} \frac{1}{\sqrt{2\omega}}\Big[ a_{\omega}^{(+)} e^{-i\omega (t+x)} +a_{\omega}^{(-)} e^{-i\omega (t-x)} + \big(a_{\omega}^{(+)}\big)^{\dagger} e^{i\omega (t+x)} + \big(a_{\omega}^{(-)}\big)^{\dagger} e^{i\omega (t-x)} \Big]\,.
\end{equation}
The above equation indicates that L-quantization describes the system in terms of massless particles, in the $x\rightarrow -\infty$ region.  
This massless particle description seems quite natural in the asymptotic left region, but it is not warranted in the whole region. As we will see in the following, the L-quantization scheme is not suitable for the right asymptotic region or the right observer.

To address this limitation, we introduce the R-quantization scheme, which would be more appropriate  for particle interpretation in the right asymptotic region, since it is based on the asymptotic form of the solution in~\eqref{phiw1}.  One may think  that  R-quantization  is equivalent to L-quantization, not alternative one, because one can rewrite mode functions $u^{(i)}_{\omega}$ in terms of another set of mode functions  using the transformations of hypergeometric functions given in \eqref{HyperP1}.  Even though the mode functions can be related through the $SU(1,1)$ transformation presented in~\eqref{transtcl}, it fails to preserve the canonical commutation relation among the operators transformed from $a_{\omega}^{(+)}$ and $a_{\omega}^{(-)}$. 
Mathematically, the condition for the equivalence of Fock spaces and the preservation of canonical commutation relations is described by the Shale theorem~\cite{Shale,Derezinski}. In our setup, the Shale theorem's condition is not met because the $SU(1,1)$ transformation is incomplete and non-invertible due to the absence of propagating modes in the right asymptotic region for the range of $0 \le \omega \le b \beta$.
After removing the unphysical exponentially increasing part among the non-propagating modes, we have the exponentially decaying parts only in the $x \to \infty$ region (as shown in \eqref{phiw1} and \eqref{phiwc}) under the range of $0\leq\omega\leq b\beta$.
Because of this reduction in modes, within the range of $ 0\leq \omega \leq b\beta$, in L-quantization scheme, $a^{(+)}_{\omega}$ and $a^{(-)}_{\omega}$ are not independent but related by the relation $a^{(-)}_{\omega} = \tilde{R}^{*}_{\omega} a^{(+)}_{\omega}$ where $R_{\omega}$ is given in~\eqref{RefC}.

We now introduce another quantization scheme, which will be referred as R-quantization. This provides a quantization of the scalar field based on the classical solution in~\eqref{phiw1} as
\begin{align} \label{phiR}
	\phi_{\rm R}(\pmb{x}) &= \int_{0}^{\infty}\frac{dk}{\sqrt{2\pi}}\frac{1}{\sqrt{2\omega}} \sum_{i=\pm}
	\bigg[ b_{k}^{(i)}v^{(i)}_{k}(\pmb{x})+\big(b_{k}^{(i)}\big)^{\dagger}\big(v^{(i)}_{k}(\pmb{x})\big)^{*} \bigg]\,,
\end{align}
where $v^{(\mp)}_{\omega}(\pmb{x})$'s are given by
\begin{align}
\!\!\!\! 	v^{(-)}_{k}(\pmb{x}) &=  (1+ e^{-bx})^{2\xi}   F\Big(A,A-C+1\,;\, A-B+1\,\Big|\, -e^{-bx} \Big) e^{-i (\omega t-kx)} \,,   \label{vm} \\
\!\!\!\!	v^{(+)}_{k}(\pmb{x}) &= (1+e^{-bx})^{2\xi} F\Big(B,B-C+1\,;\, B-A+1\,\Big|\, - e^{-bx} \Big) e^{-i (\omega t+kx)}\,. \label{vp}
\end{align}
Here,  $\omega = \sqrt{k^{2}+b^{2}\beta^{2}}$. Since the relation, $\big[F(A,A-C+1\,;\, A-B+1\,|\, - e^{-bx})\big]^{*}=F(B,B-C+1\,;\, B-A+1\,|\, - e^{-bx})$, holds, complex conjugates of mode functions are given by
\begin{align}
\!\!\!\!	\!\! 	\big(v^{(-)}_{k}((\pmb{x})\big)^{*} &=   (1+ e^{-bx})^{2\xi} F(B,B-C+1\,;\, B-A+1\,|\, - e^{-bx}) e^{i (\omega t-kx)} \,,   \label{vmcc} \\
\!\!\!\!	\!\! \big(v^{(+)}_{k}((\pmb{x})\big)^{*} &=   (1+e^{-bx})^{2\xi}   F(A,A-C+1\,;\, A-B+1\,|\, - e^{-bx}) e^{i (\omega t+kx)}\,. \label{vpcc} 
\end{align}
Like in \eqref{aCom}, the non-vanishing commutator of  creation and annihilation operators is given by
\begin{equation} \label{bCom}
[b_{k}^{(i)}, (b_{k'}^{(j)}\big)^{\dagger}]  = \delta^{ij} \delta(k -k')\,.
\end{equation}
Just like in L-quantization, one may introduce the vacuum in  R-quantization, $|0\rangle_{\rm R}$, as
\begin{equation} \label{}
b_{k}^{(\mp)}|0\rangle_{\rm R}=0\,.
\end{equation}
The Fock space constructed by $b^{(\mp)}_{k}$ and $\big(b^{(\mp)}_{k}\big)^{\dagger}$ is denoted by $\mathscr{F}_{\rm R}$.
At the right asymptotic region, $v^{(\mp)}(\pmb{x})$'s reduce to
\begin{align}
	v^{(\mp)}(\pmb{x}) \underset{x\rightarrow \infty}\longrightarrow e^{-i(\omega t\mp kx)}\,,
\end{align}
and the canonically quantized form of the scalar field at the right asymptotic region $(x\rightarrow\infty)$ reduces to
\begin{equation} \label{}
\phi_{\rm R} (\pmb{x}) \underset{x\rightarrow\infty}{\simeq} \int^{\infty}_{0} \frac{dk}{\sqrt{2\pi}} \frac{1}{\sqrt{2\omega}}\Big[ b_{k}^{(+)} e^{-i (\omega t+kx)} +b_{k}^{(-)} e^{-i (\omega t-kx)} + \big(b_{k}^{(+)}\big)^{\dagger} e^{i (\omega t+kx)} + \big(b_{k}^{(-)}\big)^{\dagger} e^{i (\omega t-kx)} \Big]\,.
\end{equation}
The above equation shows that in contrast to L-quantization, R-quantization gives rise to particles with a mass of $b\beta=m_0/a=2\xi b$, which can be inferred from $\omega=\sqrt{k^2+b^2\beta^2}$. This means that the  (local) `right' observers cannot detect massless particles of the energy, $\omega<b\beta$,
which can be observed in the `left' region.

\vskip0.3cm
\noindent Some comments are given in orders:\\
%
	We have introduced two vacua, $|0\rangle_{\rm L}$ and $|0\rangle_{\rm R}$, which are inequivalent due to the lack of an invertible transformation connecting $a^{(i)}_{\omega}$ and $b^{(i)}_{k}$. As previously mentioned, the inequivalence of the two vacua arises from the fact that the equivalence criterion stated in the Shale theorem is violated. Consequently, both quantization schemes are distinct, and neither vacuum is preferred, as is typically the case in quantum field theory in curved spacetime. In the context of algebraic approach~\cite{Haag:1992hx,Wald:1995yp}, these vacua are (local) quasi-free states satisfying the local Hadamard condition which justifies the naming  (local) ``vacua'' for $|0\rangle_{\rm L}$ and $|0\rangle_{\rm R}$. In a similar vein, it is natural to interpret the Fock spaces, $\mathscr{F}_{\rm L}$ and $\mathscr{F}_{\rm R}$, as local Hilbert spaces, rather than global ones. See for a recent review~\cite{Witten:2021jzq}.

	Even though (local)  $\mathscr{F}_{\rm L}$ and $\mathscr{F}_{\rm R}$ are not unitarily equivalent, they share the same algebraic relation among the field operators.  In this  algebraic view point, one may consider some extended (algebraic)  states from the previous local ones. Additionally, since our system exhibits time-translation symmetry, the Hamiltonian of the system, which is constructed from the field operators, can be uniquely defined. So would be  the minimum energy of the ``physical state''.\footnote{In the algebraic approach, global Hadamard states are accepted as a good candidate for ``physical states''.} Two vacua, $|0\rangle_{\rm L}$ and $|0\rangle_{\rm R}$, satisfying local Hadamard condition,\!\footnote{This means  that the two-point function's coincident limit   has a universal divergence structure given by the Minkowski case.} might be extended to global Hadamard states~\cite{Kay:1988mu,Radzikowski:1996pa}, respectively. However, these two extended vacua cannot be quasi-equivalent, since the massless particle Fock space $\mathscr{F}_{L}$ and the massive one $\mathscr{F}_{R}$ are  not unitarily  transformed into each other even if the states in those Fock spaces are purified  in some way.  Here, the concept of quasi-equivalence is characterized by the existence of a unitary equivalence between the purifications of two quasi-free states~\cite{VanDaele:1971dh,Verch:1992eg}. 
	One may argue that the global Hadamard state extended from $|0\rangle_{\rm L}$ is the true vacuum since it has a lower (renormalized) energy from an analogy of the quantum mechanical potential problem, while the one extended from $|0\rangle_{\rm R}$ would have higher energy than previous one. However, the absence of Poincar\'e invariance obstructs this conclusion. 
It is worth noting that even when $\omega > b\beta$, the operators $a_\omega$ and $b_k$ are not unitarily related. This discrepancy becomes evident in the non-preservation of the commutation relations between $(a_{\omega},a_{\omega}^{\dagger})$ and $(b_{k},b_{k}^{\dagger})$. The possibility of a unitary equivalence between $a_{\omega}$ and $b_{k}$ within the range of $\omega>b\beta$ leads to the inference that the descriptions of massless and massive particles are equivalent, a scenario that contradicts the conventional Fock space interpretation. This circumstance falls outside the scope of the valid range of the Shale theorem, which provides a mathematically rigorous framework for the equivalence of two sets of oscillators.

 Consequently, we believe that the algebraic approach would be a more superior method in this regard as advertised in~\cite{Ho:2022ibc}. We believe future works to be required in order to clarify the existence of quasi-free global Hadamard states in our case along the line of \cite{Kay:2006jn,Derezinski2018}.


%

\section{Conclusion}

There have been various sporadic studies on IFT from diverse directions. 
Compared to conventional relativistic field theory, 
IFT has numerous  unusual aspects including the difficulty in its quantization. Because of the absence of 
the Poincar\'{e} invariant vacuum, we have proposed the algebraic approach as an appropriate method for IFT quantization~\cite{Ho:2022ibc}. This proposal is motivated by a natural correspondence between FTCS and IFT in $(1+1)$ dimensions. Along this line of thought, in this paper we have explored the relation between  SFTCS and SIFT.  

After briefly reviewing the conversion between bosonic FTCS and bosonic IFT, we have presented a procedure to construct supersymmetric field theory on curved backgrounds. While this procedure is inspired by the seminal work of Festuccia and Seiberg~\cite{Festuccia:2011ws}, it differs from their method in some details. Specifically, our background is not obtained as a conventional supergravity solution.  In our approach, we have just treated our background  as a rigid one. By solving the (generalized) Killing spinor equation, we have classified possible supersymmetric cases within our setting and verified that flat and AdS$_{2}$ spaces allow two real supersymmetries. Any other supersymmetric backgrounds allow just only one real supercharge in our formulation. As a simple example, we have found a static supersymmetric background which contains a naked null curvature singularity. Interestingly, this background accommodates a wide range of supersymmetric field theories. For instance, we can incorporate scalar field theories such as $\phi^6$-theory, Liouville theory, and Sine-Gordon theory on this background.

From the viewpoint of FTCS, one may claim that a singular spacetime is not so relevant and should be excluded from consideration. However, this claim may be disputed by noting that  $(1+1)$-dimensional gravity is not dynamical in Einstein gravity. Furthermore, we showed that scalar wave propagation in our background is completely well-posed as an initial value problem.  In this regard, we have established the essential self-adjointness of the symmetric operator $A$, which is derived from the scalar wave equation. To this end, we employed the standard functional analysis technique for testing self-adjointness. Furthermore, as alluded in~\cite{Ishibashi:2004wx},  we need  case-by-case analysis for non-compact spacetimes such as AdS space.  Since our background is Lorentzian and non-compact, we provided concrete steps which reveal the essential self-adjointness of the symmetric operator $A$ and verified explicitly the global-hyperbolicity of our background spacetime from the viewpoint of scalar wave propagation. In this analysis, it is very useful to notice the role of SQM in the solvability  and its relation to scalar field theory part of free SFTCS/SIFT.

The IFT picture is very useful for comprehending and interpreting our metric background, as the behavior of massive scalar fields around the singularity in the FTCS context can be understood in terms of simple massless scalar fields on Minkowski spacetime from the IFT perspective. For instance, when $a>0$ and $\xi \ge \frac{1}{4}$ in Fig.~\ref{pa3}, we  give a concrete IFT interpretation for classical wave scattering as follows: A wave corresponding to a massless particle incident in the left region ({\it i.e.} $x\rightarrow -\infty$) with sufficient momentum (or energy) greater than $b \beta$ undergoes partial reflection and partial transmission, resulting in a wave corresponding to a massive particle  in the right region ({\it i.e.} $x\rightarrow \infty$). This process conserves total energy due to time translation symmetry, which,  viewed from the FTCS perspective, is just the standard dynamics of scalar fields with a constant mass $m_0$ on a curved background.

We would like to note that the above complementary description using both  FTCS and IFT languages is very useful for understanding relevant physics.  
Unusual-looking phenomena in one side have natural interpretations in the other side.

In addition, we have examined the quantization of the scalar field by utilizing the inter- changeable viewpoints offered by FTCS and IFT, building on our proposal~\cite{Ho:2022ibc}.
After we attempted to perform canonical quantization of  our system in two ways, namely L-quantization and R-quantization, we revealed that the canonical quantization is obscured. This is related to physical interpretation of the quantization of scalar field together with its {\it mass change}  in the IFT picture.
To address this issue, we adopted the algebraic approach to quantum field theory, which requires each Fock space to be  a local one rather than a global one.  This local interpretation of Fock spaces comes from the algebraic approach to quantum field theory, which is required to interpret the system consistently.

Though there is no essential problem in scalar wave propagation, the geodesic incompleteness of our background may ask some more consideration. 
When a wave propagates and touches the singularity within a finite affine parameter, one may wonder what happens at the singularity. 
To extend spacetime beyond the singularity, various methods may exist. We provide an example of one such method. Given the {\it mild} nature of our curvature singularity, we speculate boldly that an analogous extension method, similar to the Rindler-to-Minkowski extension,  might be used, even though the physical situations are different. 
Note that our supersymmetric background metric can be written as 
\begin{equation} \label{OurmetricA}
ds^{2} = \frac{1}{(a+e^{-bx})^{2}}(-dt^{2} + dx^{2}) = - \frac{dx^{+}dx^{-}}{(a+e^{-\frac{b}{2}(x^{+}-x^{-})})^{2}}\,. 
\end{equation}
%
By the following coordinate transformation 
\begin{equation} \label{CorTrans}
y^{+} \equiv e^{\frac{b}{2}x^{+}}\,, \qquad y^{-} \equiv - e^{-\frac{b}{2}x^{-}}\,, 
\end{equation}
one can see that the metric can be rewritten as 
\begin{equation} \label{}
ds^{2} = -\frac{-y^{+}y^{-}}{\frac{b^{2}}{4}( 1- a y^{+}y^{-} )^{2}} dy^{+}dy^{-}\,.
\end{equation}
By the analogy with the extension of the Rindler coordinates to the Minkowski ones, 
we might think the above coordinate transformation with the extended range of $y$-coordinates as the maximal extension of our background beyond the singularity. 
As mentioned previously, this example might be an interesting subject for further study.

There are various future directions we can pursue. One of them is to explore the algebraic approach at the basic level beyond the canonical quantization done in the main text. The Hadamard method and its application would be much better at least conceptually. 
For instance, it would be interesting to explore the existence of quasi-free global Hadamard state in our system. 
The integrability of various models on our background and its relation to the counter parts in AdS space would be quite interesting research subjects. 
Another important future direction is to construct higher dimensional Lorentzian supersymmetric background. 
It would also be interesting to study finite temperature phenomena and spontaneous supersymmetry breaking in the context of SFTCS/SIFT.  
As an extension of {\it mass change} of scalar field in free SIFT, it is meaningful to investigate similar effects in couplings in SIFT. Another interesting direction is to study SFTCS/SIFT in the context of the AdS/CFT correspondence.

\section*{Acknowledgments}
We appreciate conversations and discussions with Dongsu Bak, Seungjoon Hyun, Chanju Kim,   Wontae Kim, Yoonbai Kim, and Driba D. Tolla. 
This work was supported by the National Research Foundation of Korea(NRF) grant with grant numbers \\ NRF-2022R1F1A1073053, RS-2023-00249608(O.K.), NRF-2019R1A6A1A10073079(O.K. and J.H.), NRF-2019R1A2C1006639 (J.H.),  NRF-2022R1F1A1076172, NRF-2021R1F1A1062315, NRF-2020R1C1C1012330(S.-A.P.), 
 NRF-2021R1A2C1003644(S.-H.Y.) and supported by Basic Science Research Program through the NRF funded by the Ministry of Education 2020R1A6A1A03047877(S.-H.Y. and J.H.).


\newpage 

\begin{center} {\Large \bf Appendix}
\end{center}

\begin{appendix}

\section{Notation and Convention}\label{AppA}
In this Appendix, we summarize our notation and convention. Our convention of two-dimensional Clifford algebra in flat spacetime is taken as 
\begin{equation} \label{}
\{\gamma_{\rm F}^{\mu},\gamma_{\rm F}^{\nu}\}=2\eta^{\mu \nu}\,, \qquad \eta^{\mu \nu} =\text{diag}(-1,1)\,.
\end{equation}
The adjoint,  complex conjugate, and   transposed  intertwiners of gamma matrices are given by
\begin{equation} \label{}
(\gamma_{\rm F}^{\mu})^{\dagger} = -A\gamma_{\rm F}^{\mu}A^{-1}\,, \qquad  (\gamma_{\rm F}^{\mu})^{*} = B\gamma_{\rm F}^{\mu}B^{-1}\,, \qquad  (\gamma_{\rm F}^{\mu})^{T} = -C\gamma_{\rm F}^{\mu}C^{-1}\,,  \qquad C = B^{T}A\,.
\end{equation}
Anti-symmetrized gamma matrices are defined with a  normalization constant such as $\gamma_{\rm F}^{\mu \nu}\equiv\frac{1}{2}[\gamma_{\rm F}^{\mu},\gamma_{\rm F}^{\nu}]$ which satisfies the relation $(C\gamma_{\rm F}^{\mu \nu})^{T} = C\gamma_{\rm F}^{\mu \nu}$. 

By taking the explicit  representation of gamma matrices in terms of Pauli matrices $\sigma^{a}$ as $\gamma_{\rm F}^{\mu}=(i\sigma^2,\sigma^1)$, one can set 
\begin{equation} \label{}
A =  C=\sigma^{2}\,, \qquad B = {\bf 1}\,.
\end{equation}
This representation {\it i.e.} $B = {\bf 1}$ is known as Majorana representation. 
Majorana spinor $\chi$ is defined by $\bar{\chi} \equiv \chi^{\dagger}\gamma_{\rm F}^{t} = i\chi^{T}C$. Using  the symmetric property of $(C\gamma_{\rm F}^{\mu})^{T} = C\gamma_{\rm F}^{\mu}$, one can see that   $\bar{\chi}\gamma_{\rm F}^{\mu}\chi =i\chi^{T}C\gamma_{\rm F}^{\mu}\chi =0$  in this case.

Gamma matrices in curved spacetime are defined by the vielbein as $\gamma^{\mu}=e^{\mu}_{\hat{a}}\gamma^{\hat{a}}$, which lead to curved spacetime Clifford algebra as $\{\gamma^{\mu},\gamma^{\nu}\}=2g^{\mu\nu}$. Here, $\mu, \nu$ denote curved/flat spacetime indices, and the hated indices  $\hat{a}, \hat{b}$ go for tangent spacetime ones.  Covariant derivative of fermions is introduced as 
\begin{equation} \label{coderf}
\nabla_{\mu} \equiv \partial_{\mu} + \frac{1}{4}\omega^{\hat{a}\hat{b}}_{\mu} \gamma_{\hat{a}\hat{b}} \,,
\end{equation}
where $\omega^{\hat{a}\hat{b}}_{\mu}$ denotes a spin connection.  This covariant derivative reduces to $\nabla_{\mu} = \partial_{\mu} - \frac{1}{2}\omega^{\hat{t}\hat{x}}_{\mu} \sigma^{3}$ in our $(1+1)$-dimensional convention.

\section{Solving Killing Spinor Equation} \label{AppB}
In this Appendix, we solve the Killing spinor equation~\eqref{mainKS} obtained in SFTCS, which corresponds to the Killing spinor equation~\eqref{KSIFT}  in SIFT.  Before solving the Killing spinor equation, let us derive the equation~\eqref{mainKS2}.  By taking the covariant derivative on both sides~\eqref{mainKS} and antisymmetrizing the indices, we obtain 
\begin{equation} \label{}
[\nabla_{\mu},\nabla_{\nu}]\epsilon = \frac{1}{4}R_{\mu\nu}^{~~\hat{a}\hat{b}}\gamma_{\hat{a}\hat{b}} \epsilon = \nabla_{\mu}f \gamma^{\mu}\, \epsilon + f^{2}\epsilon\,.
\end{equation}
Then, by using the 2-dimensional identity of Riemann tensor and contracting $\gamma^{\hat{a}\hat{b}}$, one  can immediately deduce the equation~\eqref{mainKS2}. By using~\eqref{mainKS2}, one can verify that $f=f_{0}=constant$ implies  ${\cal R}= -\frac{1}{2}f^{2}_{0} \le 0$. Especially,  $f=0$ implies that ${\cal R}=0$.

Now, let us take the explicit component form of the supersymmetry variation parameter $\epsilon$ in~\eqref{mainKS}:
\begin{equation} \label{}
\epsilon (\pmb{x})  = e^{\frac{1}{2}\omega(\pmb{x})}%
\left( \begin{array}{l}    
\epsilon_{-}(\pmb{x}) \\ \epsilon_{+}(\pmb{x}) 
\end{array}  \right)   \,,
\end{equation}
where we use our $\gamma$-matrix convention given in Appendix~\ref{AppA}. Then, 
we obtain the component form of the Killing spinor equation as 
\begin{equation} \label{}
\partial_{\pm}\epsilon_{\pm} + \partial_{\pm}\omega\, \epsilon_{\pm}   = \pm \frac{1}{2}\, e^{\omega}\,f\,\epsilon_{\mp}\,,   
\end{equation}
By using this component form, one can easily deduce the results in~\eqref{f0=0} and~\eqref{f0ne0}.

When $\epsilon_{-} \epsilon_{+} =0$, one can take $\epsilon_{\mp} = p_{\mp}\epsilon_{0}$ with a constant Grassman variable $\epsilon_{0}$. In this case, it is straightforward to obtain 
\begin{align}    \label{peq}
\partial_{\pm}p_{\pm} + \partial_{\pm}\omega~ p_{\pm} = \pm\frac{1}{2}\,e^{\omega}\, f\, p_{\mp}\,, \qquad  \partial_{\mp}p_{\pm} =0\,.
\end{align}
Using the  additional condition~\eqref{cond3} in conjunction with $\partial _{\pm} {\cal F} = \frac{\partial  \cal F}{\partial {\cal R}}\, \partial_{\pm}{\cal R} $, one can  deduce that $p^{2}_{+}\partial_{+}{\cal R} = -p^{2}_{-}\partial_{-}{\cal R}$. When Ricci scalar ${\cal R}$ depends only on the spatial coordinate $x$, one immediately see that $p^{2}_{+} = p^{2}_{-}$ which can be taken as $p^{2}_{+} = p^{2}_{-}=1$ by a constant rescaling of $\epsilon_{0}$.  In the end,  the supersymmetry variation parameter $\epsilon$ and $f$ are given by 
\begin{equation} \label{}
\epsilon (x)   = e^{\frac{1}{2}\omega(x)} {\,\epsilon_0 \choose \pm\epsilon_0} \,, \qquad f(\mathcal{R} )
= \pm \omega' e^{-\omega}.
\end{equation}

Though we have solved the Killing spinor equation in the context of SFTCS, all the expressions can be  understood  as the corresponding ones  in SIFT   by taking $h_{\mu}=\partial_{\mu}q$ in \eqref{KSIFT} and using  \eqref{ln(x)}.

\section{A Sample Gravity Lagrangian for Our  Geometry} \label{AppC}
In this Appendix, we would like to point out that our metric solution cannot be implemented in (supersymmetrized) dilaton-Einstein gravity but in a higher derivative theory of gravity. To do this, 
let us consider a (1+1)-dimensional  specific gravity Lagrangian given by 
\begin{align}\label{SBact}
\int d^2x \sqrt{-g} \mathcal{L} = \int d^2x \sqrt{-g} \Phi \left(\mathcal{R} - \eta \nabla^2 \mathcal{R} + \eta \mathcal{R}^2 + \frac{\eta^2}{4 } \mathcal{R}^3  \right)\,,
\end{align}
where $\eta$ is a constant parameter of (length)$^{2}$ dimension. 
Equations of motion of the dilaton field $\Phi$ and metric  $g_{\mu\nu}$ are  given by 
\begin{align}\label{REOM}
0=&\mathcal{R} + \eta \mathcal{R}^2 + \frac{\eta^2}{4 } \mathcal{R}^3 - \eta \nabla^2 \mathcal{R} \,, \nonumber \\
0=&   \Big(  -\nabla_{\mu}\nabla_{\nu}+ g_{\mu\nu}\nabla^{2} \Big)\Big[\Phi\Big(1 + 2 \eta {\cal R}  + \frac{3}{4}\eta^{2} {\cal R}^{2}\Big) -\eta \nabla^{2} \Phi\Big]  +  \eta \nabla_{(\mu}\Phi \nabla_{\nu)}{\cal R}  \nonumber \\
&  +\frac{1}{2}g_{\mu\nu}  \Big[\eta (\Phi \mathcal{R}^{2}  - \nabla_{\alpha} \Phi\nabla^{\alpha} \mathcal{R} - \mathcal{R}\nabla^{2}\Phi)+  \frac{\eta^2}{2} \Phi\mathcal{R}^3  \Big]\,,
\end{align}
Then, a solution to \eqref{REOM} is given by 
\begin{align}
ds^{2}  &= \frac{1}{(a+e^{-bx})^{2}} (-dt^{2} +dx^{2}) \,, \nonumber \\
 \Phi (x) &=  \Phi_{0} \Big[ e^{-bx} +\frac{1}{2a} e^{-2bx}  -\frac{1}{24a^{3}}e^{-4bx}  + \frac{13}{240a^{4}}e^{-5bx}  -\frac{143}{2400a^{5}}e^{-6ax} +\cdots \Big] \,,
\end{align}
where $\eta = \frac{1}{a^2 b^2} $ and $\Phi_{0}$ is a constant. 
Therefore, our supersymmetric background geometry \eqref{susybackgd} would be a specific solution to the action \eqref{SBact}.

\section{Properties of Hypergeometric Function}\label{AppD}
We adopt the definition of hypergeometric function in the form of  
\begin{equation} \label{hypergeoft}
F(A,B\, ;\, C\,|\, z) = \sum_{n=0}^{\infty}\frac{(A)_{n}(B)_{n}}{(C)_{n}} \frac{z^{n}}{n!}\,, 
\end{equation}
where $(A)_{n}$ denotes the Pochhammer Symbol defined by $(A)_{n} = \frac{\Gamma(A+n)}{\Gamma(A)}$.

Let us introduce $\tilde{A}, \tilde{B}$ and  $\tilde{C}$ as 
\begin{equation} \label{}
\tilde{A} \equiv A-C+1\,, \qquad \tilde{B} \equiv B-C+1\,, \qquad \tilde{C} \equiv 2-C\,,
\end{equation}
or equivalently as 
\begin{equation} \label{}
A \equiv \tilde{A}-\tilde{C}+1\,, \qquad B \equiv \tilde{B}-\tilde{C}+1\,, \qquad C  \equiv 2-\tilde{C}\,.
\end{equation}
Then, it is straightforward to check 
\begin{equation} \label{}
\tilde{B}-\tilde{A} = B-A\,, \qquad \tilde{C}-\tilde{A}-\tilde{B} = C-A-B\,, 
\end{equation}
which lead to 
\begin{equation} \label{}
-(A+\ell) (\tilde{A}+\ell) = (\tilde{A}+\ell) (\tilde{C}-\tilde{A}-1-\ell) = (A+\ell) (C-A-1-\ell)\,.
\end{equation}
Now, one can see that
\begin{align} \label{Id1}
{\textstyle \frac{ (A)_{n} }{\Gamma(B) \Gamma(C-A-n)} } &= {\textstyle \frac{1}{\Gamma(C-A)} \prod^{n-1}_{\ell=0} A(C-A-1) \cdots (A+\ell-1)(C-A-\ell+1)} 
\nonumber\\
&= \frac{1}{R_k} {\textstyle \frac{ (\tilde{A})_{n}}{\Gamma(\tilde{B})\Gamma(\tilde{C}-\tilde{A}-n)} }\,,
\end{align}
where we used the definition of $R_{k}$ in~\eqref{RefC0} 
\begin{equation} \label{}
R_{k} = {\textstyle  \frac{\Gamma(B)\Gamma(C-A)}{\Gamma(\tilde{B})\Gamma(\tilde{C}-\tilde{A})}  }\,.
\end{equation}

The transformation rule for hypergeometric functions for $ B - A= m \notin {\bf N}$ becomes
\begin{align}    \label{HyperP1}
 \frac{\sin\pi(B-A)}{\pi\Gamma(C)} {\textstyle F(A,B\,;\, C\,|\, z) } = &  {\textstyle \frac{(-z)^{-A}}{\Gamma(B)\Gamma(C-A)\Gamma(A-B+1)}  F(A,A-C+1\,;\, A-B+1\,|\, \frac{1}{z}) } \nonumber \\
&   -  {\textstyle \frac{(-z)^{-B}}{\Gamma(A)\Gamma(C-B)\Gamma(B-A+1)} F(B,B-C+1\,;\, B-A+1\,|\, \frac{1}{z}) } \,,  
\end{align}
and for $B-A = m \in {\bf N} $, it becomes
\begin{align}    \label{HyperP2}
\frac{1}{\Gamma(C)}F(A,B\, ;\, C\,|\, z) &=\frac{(-z)^{-A}}{\Gamma(B)} \sum^{B-A-1}_{n=0}  {\textstyle \frac{(A)_{n}\Gamma(B-A-n)}{n!\Gamma(C-A-n)} } z^{-n}   \nonumber \\
& +   \frac{(-z)^{-A}}{\Gamma(A)} \sum^{\infty}_{n=0}  {\textstyle   \frac{(-1)^{n}\, (B)_{n}}{n!\Gamma(n+B-A+1)\Gamma(C-B-n)}  } z^{-n-(B-A)}   \times \Big[\ln (-z)  + \psi(n+1) \nonumber \\
&  \qquad  \qquad  \qquad  + \psi(n+B-A+1) -\psi(B+n) -\psi(C-B-n) \Big]\,,
\end{align}
where  $\psi(z) \equiv \frac{d}{dz} \ln \Gamma(z)$. 
Using the above formula and the binomial expansion of $(1+y^{-1})^{\beta}$, one can see that the leading term is given by $y^{\alpha+\beta -A}$ as
\begin{equation} \label{}
(1+y)^{\beta}y^{\alpha}F(A,B\,;\, C\,|\, -y) =  \frac{y^{\alpha+\beta -A}\Gamma(C)}{\textstyle \Gamma(B)} \sum^{\infty}_{n=0}(-1)^{n}{\textstyle \frac{\Gamma(n-\beta)}{\Gamma(-\beta)} }y^{-n}  \sum^{B-A-1}_{k=0}  {\textstyle \frac{(A)_{k}\Gamma(B-A-k)}{k!\Gamma(C-A-k)} } (-y)^{-k}  + \cdots\,.
\end{equation}

From the above formulae, coefficients $b_1$ and $b_2$ in \eqref{phiw} are related to $a_1$ and $a_2$ 
in \eqref{phisol} as follows,
\begin{align}     b_{1} &= \Gamma(B-A)\Big[ \frac{a_{1}\Gamma(C)}{\Gamma(B)\Gamma(C-A)} + \frac{a_{2}\Gamma(2-C)}{\Gamma(B-C+1)\Gamma(1-A)}\Big]\,,      \label{abrel1} \\ 
b_{2} &= \Gamma(A-B) \Big[ \frac{a_{1}\Gamma(C)}{\Gamma(A)\Gamma(C-B)} + \frac{a_{2}\Gamma(2-C)}{\Gamma(A-C+1)\Gamma(1-B)}\Big]\,.    \label{abrel2}
\end{align}
Similarly,  $c_{1,0}$ and $c_{2}$ in \eqref{phiwc} are given by 
\begin{align}    \label{}
c_{1,0} &=  \frac{\Gamma(B-A)}{\Gamma(B)\Gamma(C-A)} \,  a_{1}  +   \frac{\Gamma(B-A)}{\Gamma(B-C+1)\Gamma(1-A)} \,  a_{2}\,, \nonumber   \\ 
c_{2} &=   \frac{\Gamma(B)}{\Gamma(A)\Gamma(C-B)\Gamma(B-A+1)} \,  a_{1}  +  \frac{\Gamma(B-C+1)}{\Gamma(A-C+1)\Gamma(1-B)\Gamma(B-A+1)} \,  a_{2}  \,. \nonumber  
\end{align}
To obtain a square integrable solution,  we should impose $c_{1, n}=0$  for any $n=0,\cdots, \frac{\kappa}{b}$. 
 To this purpose, it  would be sufficient to set $c_{1,0}=0 $  to insure $c_{1,n}=0$. This can be deduced by applying  the identity in~\eqref{Id1} to~\eqref{HyperP2} in the hypergeometric functions in~\eqref{phisol}.

%
%
%

Now, let us introduce rescaled coefficients $\tilde{a}_{i}, \tilde{b}_{i}$ as\footnote{Here, $a_{i}$ and $b_{i}$ are coefficients in front of hypergeometric functions in classical solutions, \eqref{phisol} and \eqref{phiw}, respectively.} 
\begin{equation} \label{}
  \tilde{a}_{i} \equiv  a_{i}\sqrt{\omega} \,, \quad  \tilde{b}_{i} \equiv b_{i}\sqrt{k}\,,
\end{equation}
which shows us that   the coefficient relation is related by a $SU(1,1)$ transformation  
\begin{align}    \label{transtcl}
{ \tilde{b}_{1}    \choose \tilde{b}_{2}   }
= 
\left( \begin{array}{ll}    
p_{k}  & q_{k}   \\
q^{*}_{k}  & p^{*}_{k} \end{array}  \right)   
{ \tilde{a}_{1} \choose \tilde{a}_{2} }\,,    \qquad |p_{k}|^{2} - |q_{k}|^{2} =1\,,
\end{align}
where $p_{k}$ and $q_{k}$ are defined by
\begin{align}    \label{p}
p_{k} &\equiv  {\textstyle \frac{\Gamma(2i\frac{k}{b})\Gamma(1+2i\frac{\omega}{b})}{\Gamma(\beta+\frac{i}{b}(\omega+k))\Gamma(1-\beta + \frac{i}{b}(\omega+k))} \sqrt{\frac{k}{\omega} }= -ie^{\frac{i}{2} \theta_{k} }  \textstyle{ \frac{\Gamma(1+2i\frac{\omega}{b})}{\Gamma(1-2i\frac{k}{b})}} \frac{\sqrt{\sin^{2}\pi\beta + \sinh^{2}\frac{\pi}{b}(\omega+k) } }{\sinh\frac{2\pi}{b}k} \sqrt{\frac{k}{\omega}} } \,,    \\
q_{k} &\equiv {\textstyle \frac{\Gamma(2i\frac{k}{b})\Gamma(1 - 2i\frac{\omega}{b})}{\Gamma(\beta-\frac{i}{b}(\omega - k))\Gamma(1-\beta - \frac{i}{b}(\omega-k))} \sqrt{\frac{k}{\omega} }} = -ie^{-\frac{i}{2} \theta_{-k} }  \textstyle{ \frac{\Gamma(1-2i\frac{\omega}{b})}{\Gamma(1-2i\frac{k}{b})} \frac{\sqrt{\sin^{2}\pi\beta + \sinh^{2}\frac{\pi}{b}(\omega-k) }}{\sinh\frac{2\pi}{b}k}  \sqrt{\frac{k}{\omega}} }\,. \label{q}
\end{align}
Note that  $R_{k}$ is given by $R_{k} = - \frac{q^{*}_{k}}{p^{*}_{k}}$.

\section{Rindler Limit}

To consider the limit of $a\rightarrow 0$ of our differential equation or the Rindler spacetime limit, we should be careful to take an appropriate limit correctly. Since the relation $2\xi ab = m_{0}$ cannot be preserved in the limit or the relation is not compatible with the Rindler limit, we should consider a more generic  way. 
Indeed, the path we choose for the limiting procedure, going from our background to the Rindler spacetime, does not uphold the relationship $2\xi ab = m_{0}$.
In other words, we temporarily maintain the parameter $\xi$ as finite and allow $a$ to approach zero from the positive side with $m_0/b$ held at a finite value. Subsequently, we take the limit of $m_0\rightarrow0$, while regarding $m_0$ as an unphysical regulator. This approach is similar to that used in the massless scalar field theory in Rindler spacetime, where $m_0$ is also used as a regulator.

To see the limit correctly,  let us set $y  \equiv \frac{ab}{2m_{0}}u $ taking into account $a\rightarrow 0$. Then, though we can still take $\beta =\frac{m_{0}}{ab}$,   the differential equation reduces to 
\begin{equation} \label{}
\Big[ u\frac{d^{2}}{du^{2}} +  \Big( (2\alpha +1)   + u \Big)\frac{d}{du} + \frac{1}{2}(2\alpha +1) \Big]f_{\omega}(u) =0\,,
\end{equation}
whose solution is given by  a linear combination of confluent hypergeometric functions  ${}_{1}F_{1}( \alpha+\frac{1}{2}, 2\alpha +1 \,;\, -u) = M( \alpha+\frac{1}{2}, 2\alpha +1 \,;\, -u)$ and $U( \alpha+\frac{1}{2}, 2\alpha +1 \,;\, -u)$ as
\begin{align}    \label{}
f_{\omega}(u)  &=  A_{1}M\Big( \alpha+\frac{1}{2}, 2\alpha +1 \,;\, -u\Big) +  A_{2}  U\Big( \alpha+\frac{1}{2}, 2\alpha +1 \,;\, -u\Big)  \nonumber \\
&=A_{1} \Gamma(1+\alpha) e^{-\frac{u}{2}} \Big(-\frac{u}{4}\Big)^{-\alpha} I_{\alpha}\Big(-\frac{u}{2}\Big) + A_{2}\frac{1}{\sqrt{\pi}}(-u)^{-\alpha} e^{-\frac{u}{2}}K_{\alpha}\Big(-\frac{1}{2}u\Big), \nonumber  \end{align}
which can be rewritten in terms of $I_{\alpha}(u/2)$ and $K_{\alpha}(u/2)$ by using 
\begin{equation} \label{}
I_{\nu}(-z) = e^{i\pi\nu}I_{\nu}(z)\,, \qquad  
K_{\nu}(-z) = e^{-i\nu\pi} K_{\nu}(z) - i \pi I_{\nu}(z)\,.
\end{equation}
One can remove $I_{\alpha}(u/2)$ by taking a specific combination of two modified Bessel functions $I_{\alpha}(-u/2)$ and $K_{\alpha}(-u/2)$ to satisfy the boundary condition {\it i.e.} to remove the divergent behavior  as $x\rightarrow \infty$ or to obtain the complete reflection in the intermediate region. Then,  one obtains $f_{\omega}$ with a single coefficient  $\tilde{C}$  as
\begin{equation} \label{}
f_{\omega}(u) = \tilde{C}  u^{-\alpha} e^{-\frac{u}{2}} K_{\alpha}\Big(\frac{u}{2}\Big) =C  y^{-\alpha} e^{-\frac{u}{2}} K_{\alpha}\Big(\frac{u}{2}\Big)\,. \nonumber 
\end{equation}
Using 
\begin{equation} \label{}
 (1+y)^{\beta}  = \Big(1+a \frac{bu}{2m_{0}}\Big)^{\frac{1}{a}\frac{m_{0}}{b} } ~\underset{a\rightarrow 0}{\longrightarrow}~  e^{\frac{u}{2}}\,, 
\end{equation}
one can check that 
\begin{equation} \label{}
\phi_{\omega}(y) = y^{\alpha}(1+y)^{\beta}f_{\omega}(y) = C K_{i\frac{\omega}{b}}\Big(\frac{m_{0}}{b}e^{bx}\Big)\,, 
\end{equation}
which reproduces  the mode solution in the Rindler  wedge given by 
\begin{equation} \label{}
\phi_{\Omega}(t,x) \sim \frac{1}{\sqrt{2\Omega}}\theta(\rho)h_{\Omega}(\rho)e^{-i\Omega \eta}\,, \qquad \eta\equiv bt\,,\quad \rho\equiv b^{-1}e^{bx}\,.
\end{equation}
%

\section{Geodesics}\label{AppGeodesics}
Though it is rather straightforward to write down the geodesic equations for all kind of geodesics for two-dimensional metric in the conformal form
\begin{equation} \label{}
ds^{2} =e^{2\omega(\pmb{x})} (-dt^{2} + dx^{2})\,,
\end{equation}
we present some details for completeness. For simplicity, we focus on the static case only, as $\omega(\pmb{x}) =\omega(x)$. In this case, one can easily integrate the (2nd order) geodesic equations to obtain the reduced first order differential equations.

In summary, a timelike geodesic is determined by the following equation
\begin{equation} \label{}
\frac{dx}{\sqrt{1-A^{2}_{0}\, e^{2\omega(x)}}} = dt\,, 
\end{equation}
where $A^{2}_{0}$ is a non-vanishing integration constant. Then, the proper time $\tau$ is related to the coordinate time $t$ as 
\begin{equation} \label{}
d\tau =  A_{0}\, e^{2\omega(x)}dt\,.
\end{equation}
A spacelike geodesic is determined by
\begin{equation} \label{}
\sqrt{\frac{C_{0}}{e^{2\omega(x)} + C_{0}} }~ dx=dt\,, 
\end{equation}
where $C_{0}$ is another  integration constant.  The proper distance is related to the coordinate position $x$ as 
\begin{equation} \label{}
ds = \frac{e^{2\omega(x)}}{\sqrt{e^{2\omega(x)}+C_{0}}}\, dx\,. 
\end{equation}
Note that the simplest case corresponds to $C_{0}=0$ and $x$ does to the proper distance $s$. 
On the other hand,  in terms of null coordinates defined by $x^{\pm} = t\pm x$ and  their corresponding the affine parameters $\lambda$ and $\mu$,  null geodesics are determined by
\begin{equation} \label{}
d\lambda = e^{2\omega(x)} dx^{-}\Big|_{x^{+}=x^{+}_{0}}\,,  \qquad  d\mu = e^{2\omega(x)} dx^{+}\Big|_{x^{-}=x^{-}_{0}}\,, \qquad \quad x = \frac{1}{2}(x^{+}-x^{-})\,,
\end{equation}
where $x_0^{\pm}$'s are constants.

As examples, let us consider the static patch of de-Sitter space which can be represented in the above metric form by 
\begin{equation} \label{}
e^{\omega(x)} = \frac{1}{\cosh x}\,.
\end{equation}
By setting $r=\tanh x$, one may obtain a familiar form of the static patch of de-Sitter space.  In this case one can obtain the formulae for geodesics explicitly. Through the above formula, one can see that the timelike geodesic is given by
\begin{equation} \label{}
\sinh (t-t_{0}) = \frac{1}{\sqrt{1-A^{2}_{0}}}\, \sinh x\,,
\end{equation}
and the spacelike geodesic is given by 
\begin{equation} \label{}
\sinh (t-t_{0}) = \sqrt{\frac{C_{0}}{1+C_{0}}}\, \sinh x\,.
\end{equation}
For the null geodesics, one obtains
\begin{equation} \label{}
\lambda = -\frac{1}{2}\tanh\frac{1}{2}(x^{+}_{0}-x^{-})\,, \qquad  \mu = \frac{1}{2}\tanh\frac{1}{2} (x^{+} -x^{-}_{0})\,.
\end{equation}
In the  case of the Rindler patch of AdS$_{2}$ space, one obtains the equations for geodesics as 
\begin{align}    \label{}
\cosh(t-t_{0})  = \frac{1}{\sqrt{1+A^{2}_{0}}}\, \cosh x\,, & \qquad \sinh (t-t_{0}) = \sqrt{\frac{C_{0}}{1-C_{0}}}\, \cosh x\,,  \\
\lambda =  \frac{1}{2}\coth\frac{1}{2}(x^{+}_{0}-x^{-})\,,  & \qquad  \qquad     \mu = -\frac{1}{2}\coth\frac{1}{2}(x^{+}-x^{-}_{0})\,.
 \nonumber 
\end{align}

In our case, the metric and Ricci scalar are  given by 
\begin{equation} \label{Ourmetric}
e^{\omega(x)} = \frac{1}{a+e^{-bx}}\,,  \qquad R = 2ab^{2}e^{-bx}\,, 
\end{equation}
Inserting this to the generic formulae for the static case, one can convince that the spacetime described by the above metric is geodesically incomplete.

\end{appendix}

\end{document}